\documentclass[letterpaper,english,aps,prb,floatfix,showpacs,twocolumn,amsfonts,amssymb,superscriptaddress]{revtex4-2}

\usepackage[bookmarks=false]{hyperref}
\usepackage[T1]{fontenc}
\usepackage[english]{babel}
\usepackage[applemac]{inputenc}
\usepackage{graphics}
\usepackage{amssymb}
\usepackage{amsmath}
\usepackage{overpic}

\usepackage{xcolor}
\usepackage{soul}

\newcommand{\be}{\begin{equation}}
\newcommand{\ee}{\end{equation}}
\newcommand{\bea}{\begin{eqnarray}}
\newcommand{\eea}{\end{eqnarray}}

\def\a{\alpha}
\def\b{\beta}
\def\e{\varepsilon}
\def\d{\delta}
\def\g{\gamma}

\def\tt{\hat \tau}

\def\o{\omega}

\def\s{\sigma}

\def\G{\Gamma}
\def\D{\Delta}
\def\O{\Omega}

\def\pd{\partial}

\def\bk{{\bf k}}

\def\bq{{\bf q}}

\def\bQ{{\bf Q}}
\def\bA{{\bf A}}

\def\bE{{\bf E}}

\def\bJ{{\bf J}}

\def\OO{{\cal{O}}}

\def\nn{\nonumber}
\def\lb{\label}
\def\pref#1{(\ref{#1})}

\newcount\bozza \bozza=0
\ifnum\bozza=1
\newdimen\shift \shift=-2truecm
\def\lb#1{%
{\label{#1}\rlap{\kern\shift{$\scriptstyle#1$}}}}
\else\def\lb#1{\label{#1}} \fi

\begin{document}
\title{Theory of coherent-oscillations generation in THz pump-probe spectroscopy: \\ from phonons to electronic collective modes}
\author{Mattia Udina}
\affiliation{ISC-CNR and Dep.\ of Physics, Sapienza University of Rome, P.le A. Moro 5, 00185 Rome, Italy}
\author{Tommaso Cea}
\email{tommaso.cea@imdea.org}
\affiliation{IMDEA Nanoscience, C/Faraday 9, 28049 Madrid, Spain}
\affiliation{ISC-CNR and Dep.\ of Physics, Sapienza University of Rome, P.le A. Moro 5, 00185 Rome, Italy}
\author{Lara Benfatto}
\email{lara.benfatto@roma1.infn.it}
\affiliation{ISC-CNR and Dep.\ of Physics, Sapienza University of Rome, P.le A. Moro 5, 00185 Rome, Italy}

\begin{abstract}
Time-resolved spectroscopies using intense THz pulses appear as a promising tool to address collective electronic excitations in condensed matter. In particular recent experiments showed the possibility to selectively excite collective modes emerging across a phase transition, as it is the case for superconducting and charge-density-wave (CDW) systems. One possible signature of these excitations is the emergence of coherent oscillations of the differential probe field in pump-probe protocols. While the analogy with the case of phonon modes suggests that the basic underlying mechanism should be a sum-frequency stimulated Raman process, a general theoretical scheme able to describe the experiments and to define the relevant optical quantity is still lacking. Here we provide this scheme by showing that coherent oscillations as a function of the pump-probe time delay can be linked to the convolution in the frequency domain between the squared pump field and a Raman-like non-linear optical kernel. This approach is applied and discussed in few paradigmatic examples: ordinary phonons in an insulator, and collective charge and Higgs fluctuations across a superconducting and a CDW transition. Our results not only account very well for the existing experimental data in a wide variety of systems, but they also offer an useful perspective to design future experiments in emerging materials. 
\end{abstract}
\date{\today}

\maketitle
In the last decade, a significant advance in the investigation of complex systems has been gained thanks to the huge experimental progress in time-resolved spectroscopic techniques\cite{giannetti_review,cavalleri_review}. On very general grounds, the basic idea behind any pump-probe protocol is to first excite the system with a short and very intense electromagnetic pulse (pump), and then to monitor its relaxation towards equilibrium by using a secondary, weak pulse (probe) applied with a finite time delay with respect to the pump. This general protocol can then be implemented in several different ways, according to the nature of the spectroscopic measurement (angle-resolved photoemission, optical reflection or transmission, etc.) or to the wavelength of the pump/probe fields. However, in all cases one has to face with two phenomena which mark the difference with respect to ordinary equilibrium spectroscopies: (i) the use of an intense pulse triggers in general non-linear optical processes; (ii) the subsequent relaxation encodes by definition non-equilibrium phenomena on time scales which depend on the characteristics of the experiment and of the system under investigation. Due in part to these innovative aspects, many pump-probe protocols still lack a general theoretical framework able to connect the measured quantities to the material properties. More specifically, while Kubo linear-response theory\cite{mahan} represents nowadays the standard theoretical tool needed to compute the optical response of any system to a weak external perturbation, an analogous protocol for time-resolved spectroscopies has not been established yet. 

The present work aims at filling in part this knowledge gap by providing a general theoretical scheme to understand the origin of time-resolved oscillations observed in typical THz pump-THz/eV probe measurements. This experimental technique has been successfully applied in the last years to investigate the properties of both conventional superconducting (SC) compounds\cite{shimano_prl12,shimano_prl13,shimano_science14,giorgianni_natphys19,wang_natphot2019,shimano_review19}, whose spectral gap $\Delta$ lies in the THz region, and layered superconductors\cite{cavalleri_natphys16,cavalleri_science18} with an interlayer Josephson plasma frequency again in the THz range. The general scheme consists in probing the effects of the intense THz pulse by measuring the pump-induced changes $\delta E_{probe}(t_{pp})$  in the transmitted (for a thin film) or reflected (for a bulk crystal) probe field at a fixed observation time, as a function of the pump-probe delay $t_{pp}$. The observed signal is seen to oscillate at frequencies that match the typical energy scale of collective electronic excitations of the broken-symmetry state, becoming a preferential tool for investigating its nature. On the other hand, the same protocol can also be used to excite collective modes of the underlying lattice, as shown e.g.\ by the marked oscillations of the transmitted electric field at the characteristic frequency of Raman-active phonons in insulators\cite{kampfrath_prl17,johnson_prl19}. These experiments with THz pulses have their counterpart in time-resolved spectroscopies using pump and probe fields in the visible (VIS) or infrared (IR) range. Also in this case persistent oscillations of the probe field (usually collected in reflection configuration) as a function of the pump-probe time delay lie in the THz range, and they have been ascribed either to collective electronic modes, like in the case of superconducting\cite{carbone_pnas12} or charge-density\cite{yusupov_natphys10} ground states, or to Raman-active phonons (see e.g.\ Ref.\ \onlinecite{merlin_SSC97} and references therein). 

So far, the general understanding of these experiments followed different routes in the literature. For phonons, time-resolved oscillations at their typical frequencies induced by a pump in the visible light have been well understood as an impulsive-stimulated Raman scattering process (ISRS). In this case the pump field generates a coherent ion motion thanks to a difference-frequency process (DFP) analogous to stimulated Raman, where the mismatch in the energy of two optical photons can excite a phonon in the THz range. At a second stage, the change induced by the ion motion in the refractive index of the medium is detected by the probe field. This picture\cite{merlin_SSC97} has been successfully applied also to explain the difference between displacive and impulsive generation of coherent ion motion in opaque or transparent media, respectively. For what concerns instead the oscillations induced by THz pump fields, a similar scheme has been recently used to explain the experiments in wide-gap insulators like diamond\cite{kampfrath_prl17,maehrlein_prb18} and CdWO$_4$\cite{johnson_prl19}, with the remarkable difference that in this case the generation of a coherent ion motion by the pump field is made possible by a Raman-like two-photon sum-frequency process\cite{maehrlein_prb18} (SFP).  

For what concerns collective excitations in broken-symmetry states, the theoretical work has followed initially a slightly different path. As an example, in the case of superconducting systems the main focus has been put on the effects of the pump on the pairing interaction itself, more than on the overall description of the electromagnetic pump-probe process. By then assuming that the pump field induces an instantaneous quench of the pairing interaction, the SC order parameter displays, while relaxing back to equilibrium, marked oscillations at the frequency $2\Delta$ corresponding to twice the SC gap\cite{volkov73}. This perspective, along with the explicit observation of the $2\Delta$ oscillations in conventional\cite{shimano_prl12,shimano_prl13} and unconventional\cite{carbone_pnas12} superconductors,  stimulated extensive theoretical work focusing on the time relaxation of the amplitude of the order-parameter, also named Higgs mode\cite{shimano_review19}, in the presence of a time-dependent pairing interaction\cite{spivak_prl04,werner_prb16,millis_prb17,lorenzana_prb18}.  In this approach the electromagnetic (e.m.) field does not appear explicitly, since its effect is assumed to be captured by a specific time dependence of the pairing interaction. One step forward towards a microscopic description has been developed instead in Refs  \onlinecite{axt_prb07,axt_prb08,carr_sust13,manske_prb14, manske_natcomm16,kollath_prl17,manske_cm17,kemper_cm19}, where the coupling of the e.m.\ field to the electronic degrees of freedom is explicitly included. This generates a quadratic coupling of the field to the SC order parameter, that acts as a forcing term for the Higgs oscillations. In this context there have been also several attempts\cite{axt_prb08,manske_prb08,carr_sust13,manske_prb14,lenarcic_prb14,orenstein_prb15,manske_natcomm16,shao_prb16,kemper_cm19,bittner_jpsj19}, within different approximations, to define a time-dependent optical conductivity, that is however not the same quantity as the differential probe field $\delta E_{probe}(t_{pp})$ measured in Refs \onlinecite{shimano_prl13,shimano_science14,giorgianni_natphys19,wang_natphot2019}. What is to some extent missing in this description is the general  explanation of why the order-parameter fluctuations will cause an oscillation of the transmitted field as a function of the pump-probe delay, and under which conditions (i.e.\ central energy and width of the pump-field spectrum) they will be more easily detected.

An overall theoretical description of the full pump-probe process in the transmission geometry has been recently attempted in Ref.\ \onlinecite{giorgianni_natphys19} and applied to the investigation of the so-called Leggett mode, which measures the relative fluctuations of the SC phase between two bands in multi-band superconductor MgB$_2$. In this THz pump-THz probe experiment, the relative change  $\delta E_{probe}(t_{pp})$ in the transmitted probe field with and without the pump has been explicitly computed and compared to the experiments.  This allowed the authors to link the observed oscillations to the Leggett mode, which contributes in MgB$_2$ to the Raman kernel, and to show that  the basic underlying excitation mechanism is a SFP, in full correspondence with the case of phonons\cite{kampfrath_prl17,maehrlein_prb18}. 

In this manuscript we show that an analogous theoretical scheme can be formulated in full generality, allowing one to understand the pump-probe detection of collective bosonic modes in several systems. The possible applications range from ordinary phonons to collective modes across a phase transitions, like the soft phonon mode in CDW systems or the collective charge and order-parameter $2\Delta$ oscillations in superconductors. We will also establish a close correspondence between pump-probe measurements and ordinary transmission experiments at high fields, able to detect higher harmonics of the incoming radiation. This connection is particularly important for SC systems, where recent experiments with multicycle narrowband pulses\cite{shimano_science14,cavalleri_science18,keiser_cm19,wang_natphot2019} clearly established that below $T_c$ a sizable third-harmonic generation (THG) emerges. It will be shown that all these effects are strongly interconnected, and can be theoretically ascribed to the existence of a marked resonance at a certain energy scale $\omega_{res}$ in the non-linear optical kernel $K(\omega)$. Such resonance is responsible for the THG in transmission measurements as well as for the differential probe-field oscillations in pump-probe experiments, both due to a SFP induced by the pump field. In the case of THG, its intensity will then be maximized when twice the pump frequency matches the resonance value, i.e. $2\Omega_{pump}=\omega_{res}$. For what concerns instead pump-probe experiments, the oscillation frequency $\o_{osc}$ depends on the convolution between $K(\o)$ and the power spectrum of the squared pump field, consistently with a SFP. As a consequence, for broadband pulses the spectral components of $\delta E_{probe}(t_{pp})$ are dominated by the characteristic frequency of the mode, $\o_{osc}=\o_{res}$, while for narrowband pulses $\delta E_{probe}(t_{pp})$ oscillates at twice the pump frequency, $\o_{osc}=2\Omega_{pump}$, with an amplitude $K(2\Omega_{pump})$ maximized when $2\Omega_{pump}=\omega_{res}$. These expectations match rather well the observations reported so far in several SC systems\cite{shimano_prl12,shimano_prl13,shimano_science14,giorgianni_natphys19,wang_natphot2019}. From the theoretical point of view, establishing such a general paradigm provides also a scheme for further theoretical investigations aimed at linking the experimental observations to the properties of the underlying microscopic model. On this respect, the recent experimental\cite{shimano_science14,shimano_prb17,cavalleri_science18,keiser_cm19} and theoretical\cite{shimano_science14,aoki_prb15,cea_prb16,aoki_prb16,cea_leggett_prb16,shimano_prb17,cea_prb18,shimano_cm19,silaev_cm19} work focusing on the THG in superconductors, and the ongoing discussion on the nature of the collective mode responsible for it\cite{shimano_review19}, becomes crucial also for understanding the $\delta E_{probe}(t_{pp})$ oscillations in pump-probe experiments, since both are linked to the same $\omega_{res}$ resonance in the non-linear optical kernel. This link is also particularly interesting from an experimental point of view, considering the huge potential for high-harmonic generation measurements made possible nowadays by the existence of high-intensity coherent THz sources, like e.g.\ the TELBE beam-line at HZDR\cite{telbe}.

The paper is organized as follows. In Sec.\ I we introduce the general formalism needed to understand the pump-probe protocol, and we establish the general connection between the measured quantities and the microscopically computed non-linear optical kernel. The applications to few paradigmatic cases is then developed in the subsequent Sections. In Sec.\ II we first analyze the case of ordinary phonons, and we establish the full analogy between our scheme and the description given so far in the literature. In Sec.\ III we move the focus to electronic collective excitations in ordinary superconductors, where the non-linear optical kernel is dominated by a resonance at $2\Delta$. In Sec.\ IV we describe the peculiar case of the soft-phonon mode in CDW systems, where electronic and ionic excitations are strongly interconnected. Sec.\ V  is devoted to concluding remarks and discussion. 

\section{Schematic of the pump-probe experimental setup and theoretical description}
\label{general-theory}

\begin{figure}
\centering{
\hspace{-0.4cm}
\includegraphics[scale=0.4]{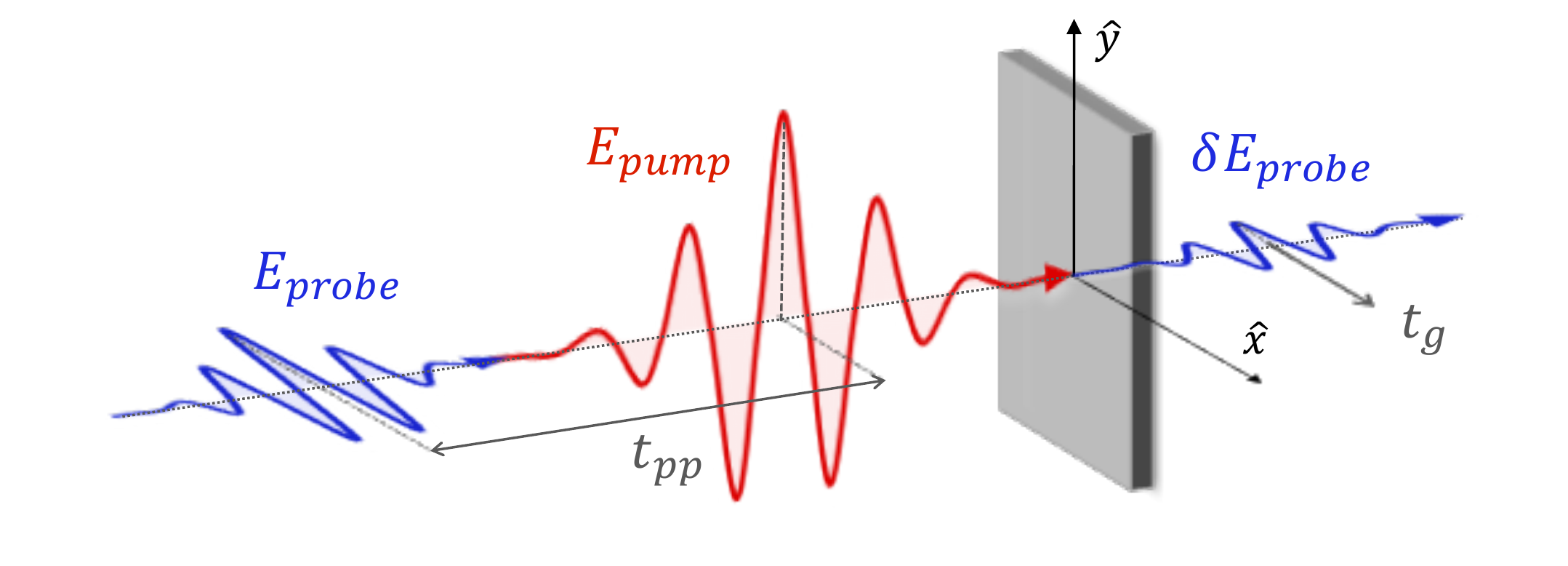}}
\caption{Scheme of the experimental configuration for the pump-probe protocol used e.g.\ in Ref.\  \onlinecite{shimano_prl13,shimano_science14,giorgianni_natphys19}. 
The pump and the probe fields are both orthogonally incident on the sample, but they are polarized in two perpendicular directions with respect to the sample surface. The transmitted probe field is then recorded for a fixed observation time $t_g$ as a function of the delay $t_{pp}$ between the two pulses.
}\label{schematic}
\end{figure}

Even though most of the present discussion is valid for pump-probe experiments done both in transmission and reflection geometry, for concreteness we will  focus on the transmission configuration, as used e.g.\ in Ref.\ \onlinecite{shimano_prl13,shimano_science14,giorgianni_natphys19}.
The protocol consists in first perturbing the sample with an intense pump field $E_{pump}$, then probing it with a  weaker probe field $E_{probe}$ and finally measuring the transmittance of the probe after the sample, as schematically depicted in Fig.\ \ref{schematic}. The observable quantity is the change in the transmitted field $\delta E_{probe}(t_{pp})$  with and without the pump, recorded as a function of the time-delay $t_{pp}$ between the pump and probe pulses. To better isolate the effects of the pump field, its polarization is usually taken orthogonal to that of the probe. For experimental reasons, it is often convenient to study the time evolution of the transmitted probe field by fixing the observation time at $t = t_{gate}$ while changing the time delay $t_{pp}$ between the pump and the probe signal (for further technical details see Refs \onlinecite{shimano_prl13,shimano_science14,giorgianni_natphys19}).  

In order to reproduce the experimental configuration, let us define the incident electric field as:
\be\lb{Ein}
\mathbf{E}_{in}(t)=E_{probe}(t)\hat{x}+E_{pump}(t)\hat{y},
\ee
with $x,y$ the crystallographic directions of the sample. By recalling the theory of transmission from thin films\cite{felderhof,katayama,minami}, the transmitted probe field can be written as
\be\lb{Etr}
E^{tr}_x(t)=\frac{1}{n+1}\left[2E_{probe}(t)-Z_0dJ_x(t)\right]
\ee
where $n$, $Z_0$ and $d$ are the refractive index, the impedance of the free space and the thickness of the film, respectively. 
$J_x$ is the $x$-component of the electromagnetic current $\mathbf{J}$ flowing inside the sample as the response to the incident field, that can be decomposed into a linear contribution and a non-linear one: $\mathbf{J}=\mathbf{J}^L+\mathbf{J}^{NL}$. For isotropic systems the linear response is diagonal in the space indices, meaning that in the field configuration \pref{Ein} it can be written as:
\be\lb{JL}
J_x^L(t)=\int\,dt'\sigma(t-t')E_{probe}(t'),
\ee 
where $\sigma$ is the optical conductivity. In the case of a phase transition, one should in principle take care about the changes in the optical conductivity across the critical temperature $T_c$.  However, as we will show below, the linear current does not give any contribution to the differential transmitted field, so we will not need to specify its behavior in the following. For anisotropic systems, in particular for layered materials, the field configuration will depend on the collective excitation one is interested in. For example, for layered cuprates to probe superconducting amplitude or density modes one should polarize the electric field in the plane, as done indeed in the recent measurement of THG in Ref.\ \cite{shimano_cm19}. On the other hand when the electric field is polarized perpendicularly to the layers one can measure the response of the out-of-plane plasma mode,  as shown in Refs\ \cite{cavalleri_natphys16,cavalleri_science18}.

The non-linear current has in principle a contribution both from the pump and the probe field, even though one already expects the latter to be predominant since the pump field is more intense. The derivation of the non-linear current starting from a given microscopic model is a conceptually clear problem, even though in practice its calculation can be non trivial for interacting systems and in the presence of a phase transition. Here we discuss the problem within the field-theory approach developed e.g.\ in Refs \onlinecite{cea_prb16,cea_prb18} for the case of the SC transition, and we apply this formalism to different physical situations. By starting from a suitable microscopic electronic Hamiltonian $H$, the gauge field ${\bf A}$ is introduced via the minimal-coupling substitution. Once fermionic degrees of freedom have been integrated out, one ends up with an effective action $S(\bA)$ for the gauge field only. Since in full generality the current is defined as the derivative of $S(\bA)$ with respect to $\bA$, we then get:
\be
\lb{jgen}
\bJ=-\frac{\pd S}{\pd \bA}, \quad \bJ^L=-\frac{\pd S^{(2)}}{\pd \bA} \quad \bJ^{NL}=-\frac{\pd S^{(4)}}{\pd \bA},
\ee
where $S^{(n)}$ is the term in the effective action containing the $n^{th}$ power of the gauge field $\bA$, whose coefficients $K^n$ are electronic susceptibilities of order $n-1$.  Therefore, the linear current is connected to the usual electronic susceptibility computed in linear-response theory, leading to the standard definition \pref{JL}, while the non-linear current is controlled by a third-order tensor $K_{\a\b\g\d}$ which depends, in general, on four spatial indexes and three incoming frequencies when $S^{(4)}$ is expressed in the frequency domain:
\begin{widetext}
\begin{equation}
\lb{s4gen}
S^{(4)}(\bA)=e^4 \int \left(\prod_{i=1}^3d\O_i \right)A_\a(\O_1)A_\b(\O_2)\tilde K_{\a\b\g\d}(\O_1,\O_2,\O_3)A_\g(\O_3)A_\d(-\O_1-\O_2-\O_3),
\end{equation}
\end{widetext}
whit $e$ the electron charge. 
While Eq.\ \pref{s4gen} accounts for all the  possible third-order processes contributing to the current, here we argue that the oscillations in $\delta E_{probe}(t_{pp})$ are present when $S^{(4)}$ admits a term that can be written as
\bea
S^{(res)}(\bA)&=&\int d\O A^2_{\a\b}(\O)K_{\a\b\g\d}(\O)A^2_{\g\d}(-\O)=\nn\\
\lb{s4res}
&=&\int dt dt' A^2_{\a\b}(t)K_{\a\b\g\d}(t-t')A^2_{\g\d}(t'),
\eea
where $A^2_{\a\b}(\O)=\int d\o A_\a(\O)A_\b(\O-\o)$ is the Fourier transform of $A_\a(t)A_\b(t)$ and the kernel $K(\O)$ displays a resonance at a given frequency $\omega_{res}$.
Let us then focus on the contributions to the non-linear current following Eq.\ \pref{s4res}, by taking into account that in an isotropic system we expect only some components of the non-linear tensor to be different from zero. As an example, for a square lattice one finds that only $K_{\a\a;\a\a}, K_{\a\a;\b\b}$ and $K_{\a\b;\a\b}$  are non zero. In the cross-polarized configuration discussed here, the incoming field \pref{Ein} has both an $x$ and $y$ component, but what matters for the transmitted field $E^{tr}$ is only its component along $x$, i.e.\ along the direction of the probe. By using Eq.s \pref{jgen} and \pref{s4res} it can be immediately shown that the $x$-component of the non-linear current is given by:
\bea\lb{JNL}
J^{NL}_x(t)&=&-2e^4A_{probe}(t)\int\,dt'\left\{K_{xx;xx}(t-t')\left[A_{probe}(t')\right]^2\right.\nn\\
&+&\left.K_{xx;yy}(t-t')\left[A_{pump}(t')\right]^2\right\},
\eea 
where $A_{probe}$ $\left(A_{pump}\right)$ is the incident vector potential of the probe (pump) field. 
For vanishing time delay $t_{pp}\simeq 0$ additional terms can be present in the non-linear response. One contribution is the  so-called instantaneous response, due to time-independent components of the non-linear kernel\cite{cea_prb18,giorgianni_natphys19}:
$$
J^{NL,inst}_x(t)=-2e n^{el}_{xx;yy} A_{probe}(t) A_{pump}^2(t) 
$$
where $n^{el}_{xx;yy}=\sum_{\bk,a}\pd^4 \e^a_{\bk}/\pd k_x^2\pd k_y^2$ and $\e_\bk^a$ is the electronic band dispersion of a given band $a$. This contribution, that is analogous to the diamagnetic response for usual linear response, can be important for the THG in transmission experiments\cite{cea_prb18}, but it does not contribute to oscillations as a function of $t_{pp}$ at large time delay. Analogously, in the presence of a finite $K_{xy;xy}$ kernel the current admits also a contribution similar to Eq.\ \pref{JNL} where the integral contains the product $A_{pump}(t')A_{probe}(t')$. However, apart from the transient regime $t_{pp}\simeq 0$ when the pump and probe overlap, this term vanishes, so it cannot contribute to the long-delay oscillations we are interested in.  Consistently with the usual analysis of the experimental data\cite{shimano_prl13,cavalleri_natphys16}, all the contributions to the early-time response will be discarded in the following. As mentioned above, one usually monitors the variation $\delta E_{probe}(t)$ in presence and in absence of the pump\cite{shimano_prl12,shimano_prl13,shimano_science14,giorgianni_natphys19}. By then using Eq.s \pref{Etr}, \pref{JL} and \pref{JNL} we find that all the contributions (linear and non-linear) proportional only to the incident probe field cancel out, so that  $\delta E_{probe}(t)$ can be expressed in terms of the non-linear response to the pump field only:
\be\lb{deltaE_probe}
\delta E_{probe}(t)=\alpha A_{probe}(t)\int\,dt'K_{xx;yy}(t-t')\left[A_{pump}(t')\right]^2,
\ee
where $\alpha \equiv \frac{2Z_0de^4}{n+1}$. In the experiments the transmitted field is recorderd at a fixed $t=t_g$, and the probe field comes with a time delay $t_{pp}$ with respect to the pump. By then rescaling $A_{pump}(t)=\bar{A}_{pump}(t+t_{pp})$, where we assume that the time profiles of both $A_{probe}(t)$ and $\bar{A}_{pump}(t)$ are centered around $t=0$, we can explicitly rewrite Eq. \pref{deltaE_probe} as:
%
\bea\lb{deltaE_tpp}
& \delta E_{probe}(t=t_g;t_{pp})= \alpha A_{probe}(t_g)\times\nn\\
& \times \int\,dt'K_{xx;yy}(t_g+t_{pp}-t')\left[\bar A_{pump}(t')\right]^2.
\eea
%
The previous equation shows that when $t_g$ is fixed the differential transmitted field becomes a function of the time delay $t_{pp}$ only. The probe field appearing in the rhs of Eq.\ \pref{deltaE_tpp} acts only as a multiplying factor, fixing the overall amplitude and phase of the oscillations. More interestingly, if we Fourier transform Eq.\ \pref{deltaE_tpp} with respect to $t_{pp}$ we find a very simple expression  for the power spectrum of the transmitted differential field:
\be\lb{E_omega}
\delta E_{probe}(\o)\propto K_{xx;yy}(\omega)\bar{A}_{pump}^2(\omega).
\ee

\begin{figure*}[t]
\centering
\includegraphics[scale=0.5]{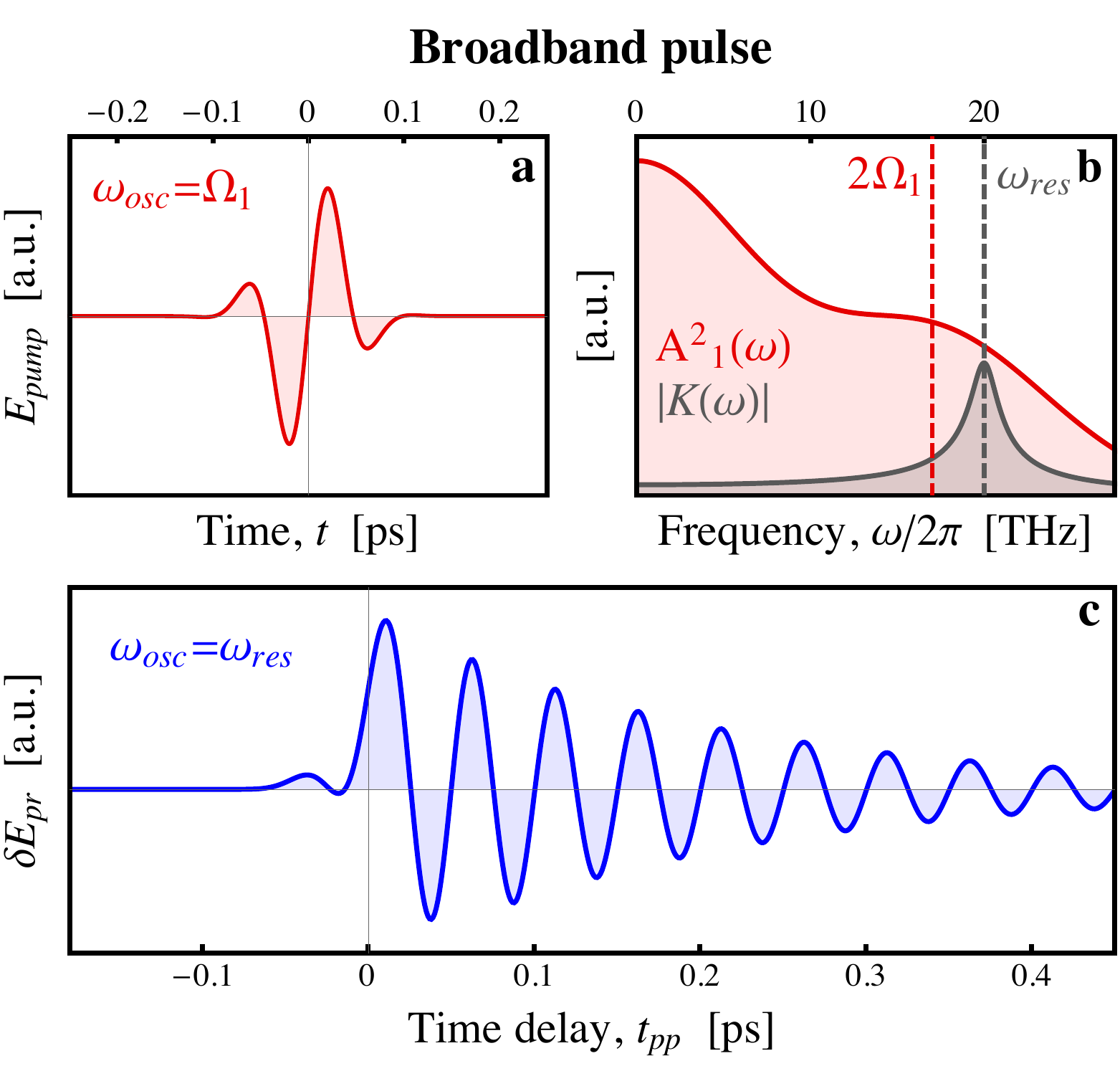}
\hspace{0.8cm}
\includegraphics[scale=0.5]{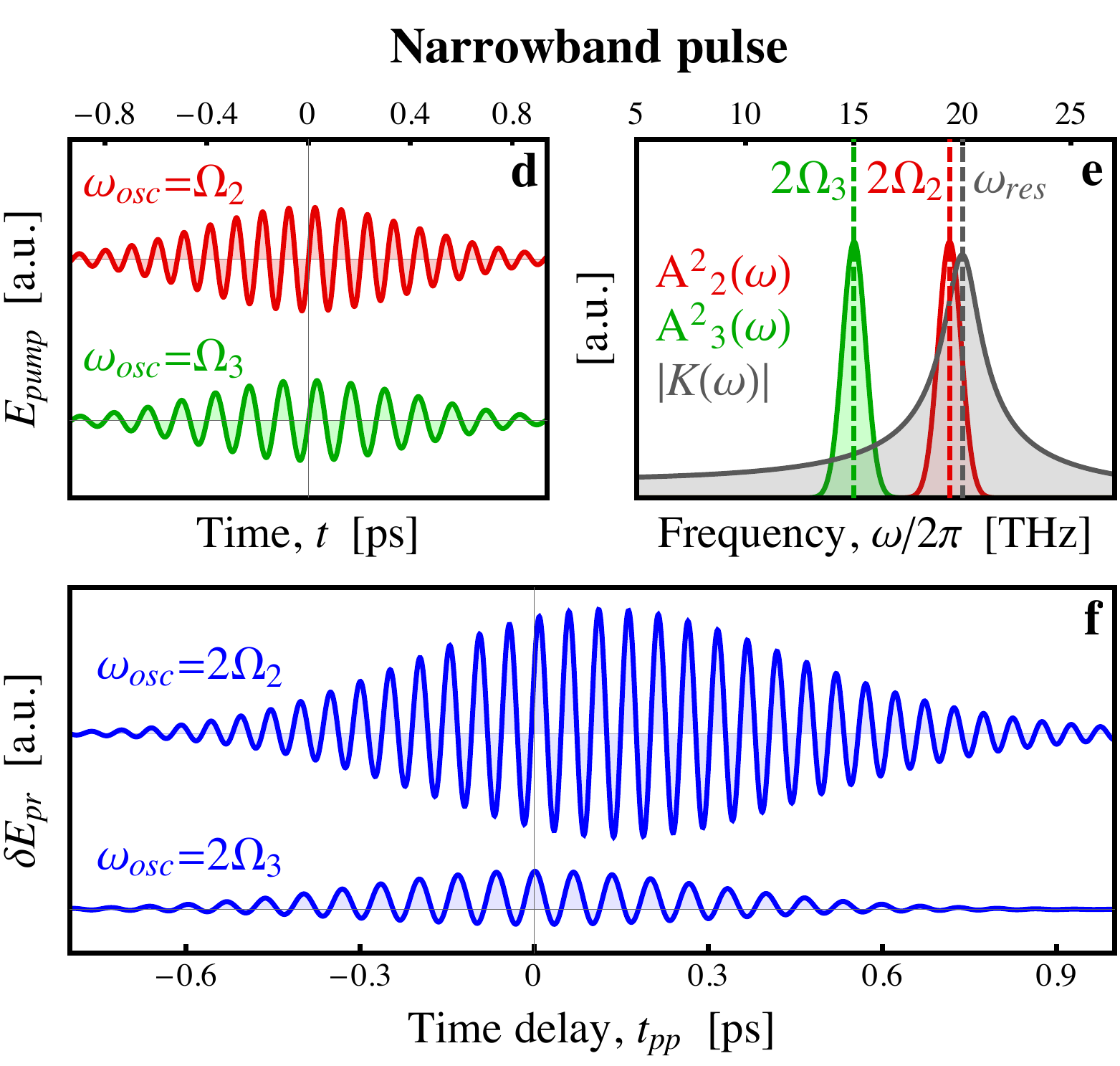}
\caption{Schematic of the mechanism responsible for the oscillations in $\delta E_{probe}(t_{pp})$. Left: broadband pulse. When the pump duration is much shorter than the mode lifetime, panel \textbf{a}, the Fourier transform of the pump field squared $A^2(\omega)$ is approximately constant around the mode resonance at $\omega_{res}$, panel \textbf{b}. As a consequence the convolution \pref{E_omega} between the pump and the mode is peaked at $\omega_{res}$, which identifies the frequency of the oscillations of $\delta E_{probe}(t_{pp})$ in time domain, panel \textbf{c}. Right: broadband pulse. In this case a longer pump pulse, panel \textbf{d}, gives rise to a sharp $A^2(\omega)$ in the frequency domain, which can be even narrower than the mode itself, panel \textbf{e}. When this condition is satisfied, the convolution $\delta E_{probe}(\omega)$ is peaked at $2\Omega_{pump}$, which sets the frequency of the  $\delta E_{probe}(t_{pp})$ oscillations. If the pump frequency matches the resonance condition $2\Omega_{pump}\approx \omega_{res}$ the amplitude of the oscillations is strongly enhanced, panel \textbf{f}. Here $t_g=0$ has been set in all the simulations.}
\label{fig-scheme}
\end{figure*}

Notice that the simple Eq.s \pref{deltaE_tpp} and \pref{E_omega} contain all the basic ingredients needed to describe resonant oscillations via a pump-probe experiment, and they allow us to distinguish the response to a broadband or a narrowband pulse in a straightforward way. To fix the ideas, let us consider the case of a Lorentzian-like resonance (as in the case e.g.\ of phonons) with central frequency $\omega_{res}$ and half-width at half-maximum $\gamma_{res}$, as it follows from a kernel
\be
\lb{ksharp}
K(\o)=\frac{1}{\o^2-\omega^2_{res}+i\omega\gamma_{res}},
\ee
where we drop for simplicity the spatial indexes of the non-linear tensor. In full generality, and for the sake of explicit calculations, we can model the incoming pulses as periodic oscillations convoluted with a gaussian decay\cite{maske_epl13,cea_prb16}, i.e.
\bea
A_{probe}(t)&=&A^0_{probe}e^{-\left (\frac{2t\sqrt{\ln2}}{\tau_{probe}}\right)^2}\cos(\Omega_{probe}t)\\
\lb{eqa}
\bar{A}_{pump}(t)&=&A^0_{pump}e^{-\left (\frac{2t\sqrt{\ln2}}{\tau_{pump}}\right)^2}\cos(\Omega_{pump}t),
\eea
where $\tau_{pump/probe}$ is the full width at half maximum, setting the pulse duration, and $\Omega$ its central frequency. The Fourier transform $\bar{A}^2_{pump}(\omega)$ appearing in Eq.\ \pref{E_omega} is then given by:
\bea
&\bar{A}^2_{pump}(\omega)=\frac{\left(A^0_{pump}\right)^2\sqrt{\pi}}{4}\frac{\tau_{pump}}{2\sqrt{\ln2}}
\left\{ e^{-\frac{1}{8}\left[\frac{(\omega-2\Omega_{pump})\tau_{pump}}{2\sqrt{\ln2}}\right]^2} +\right.\nn\\
\lb{asquared}
&+\left. e^{-\frac{1}{8}\left[\frac{(\omega+2\Omega_{pump})\tau_{pump}}{2\sqrt{\ln2}}\right]^2}+
2e^{-\frac{1}{8}\left(\frac{\omega\tau_{pump}}{2\sqrt{\ln2}}\right)^2}
 \right\},
\eea
which displays peaks at frequencies $\omega=0,\pm2\Omega_{pump}$, with approximate width $\gamma_{pump}=2\ln2/\tau_{pump}$. From Eq.\ \pref{E_omega} one immediately sees that what matters is the overlap of $A_{pump}^2(\omega)$ with $K(\omega)$ at $\omega\simeq\omega_{res}$,  with the resonance frequency $\omega_{res}$ in the THz range. This already explains the difference between experiments performed with a pump pulse centered in the visible or in the THz. In the former case $2\Omega_{pump}\gg \omega_{res}$, so what matters in Eq.\ \pref{E_omega} is only the $\omega=0$ peak in $A_{pump}^2(\omega)$. From the point of view of the photons involved in the process, this is equivalent to say that the mode is excited via a difference-frequency process where the two photons involved have energies $\omega\simeq \O_{pump}$ and $\omega\simeq -\O_{pump}+\omega_{res}$. Moreover, since pulses in the IR-VIS usually have a $\tau_{pump}\sim 10$ fs, one immediately realizes that the peak around $\omega=0$ reduces to a constant value in the THz range. On the other hand, for pulses centered in the THz range, where $\O_{pump}\simeq 0.1- 10$ THz and $\tau_{pump}\simeq 0.1-10$ ps, the overlap with $K(\omega)$ in Eq.\ \pref{E_omega} is usually made possible by the  $\omega=2\O_{pump}$ peak in $A^2_{pump}$. As a consequence, in this case two photons in the pump field with similar energies $\omega\simeq \omega_{res}$ must be added up to excite the collective mode, meaning that one is realizing a sum-frequency excitation process\cite{kampfrath_prl17,maehrlein_prb18,giorgianni_natphys19}. 

Once established the general mechanism, to determine the form of the $t_{pp}$ oscillations one should compare the relative duration of the pump pulse to the mode lifetime $\tau_{res}$, related to the width $\gamma_{res}=1/\tau_{res}$ of the $\omega_{res}$ resonance in the non-linear kernel $K(\omega)$. Indeed, from Eq.\ \pref{E_omega} one easily understands that in the so-called antiadiabatic regime where the time duration of the pulse is much shorter than the mode lifetime, i.e.\ $\gamma_{pump}\gg\gamma_{res}$ (broadband pulse), also for a THz pulse $A_{pump}^2(\omega)$ is rather flat around the maximum $\omega_{res}$ of the non-linear kernel, see Fig.\ \ref{fig-scheme}a-b. It then follows that the main frequency dependence of $\delta E_{probe}$ is determined by the optical kernel, i.e.\ $\delta E_{probe}(\omega)\sim K(\omega)$, meaning that its Fourier transform in the $t_{pp}$ time domain coincides with the Fourier transform of Eq.\ \pref{ksharp}:
\begin{widetext}
\be
\lb{broad}
\delta E_{probe}(t_{pp})\propto A_{probe}(t_g)\sin (\omega_{res}(t_{pp}+t_g))e^{-\gamma_{res}(t_{pp}+t_g)}\theta(t_{pp}+t_g),
\ee
\end{widetext}
where $\theta(t)$ is the theta function. One then finds that for a (monocycle) broadband pulse $\delta E_{probe}$ oscillates with a frequency $\omega_{osc}=\omega_{res}$, see Fig.\ \ref{fig-scheme}c. When instead the system is excited with (multicycle) narrow-band pulses (Fig.\ \ref{fig-scheme}d), i.e.\ $\gamma_{pump}\ll \gamma_{res}$, the opposite situation occurs and the convolution between the pump and the kernel in Eq.\ \pref{E_omega} can be approximated with a delta-like signal at $\omega=2\O_{pump}$, with a prefactor given by $K(2\Omega_{pump})$, see Fig.\ \ref{fig-scheme}e. In this case one expects to recover oscillations at $\omega_{osc}=2\O_{pump}$:
\begin{widetext}
\be
\lb{narrow}
\delta E_{probe}(t_{pp})\propto A_{probe}(t_g)|K(2\Omega_{pump})|\sin (2\Omega_{pump}(t_{pp}+t_g))e^{-2\left(\frac{2\ln 2(t_{pp}+t_g)}{\tau_{pump}} \right)^2},
\ee
\end{widetext}
with an amplitude that gets strongly enhanced when $\O_{pump}=\omega_{res}/2$, as depicted in Fig.\ \ref{fig-scheme}f. 
It is worth noting that the change of the frequency oscillations when going from broadband to narrowband pulses has been experimentally seen for SC resonant modes\cite{shimano_prl13,shimano_science14,giorgianni_natphys19,wang_natphot2019}. In this case the long-time decay of the measured signal can show different features according to the specific properties of the non-linear optical kernel, as we will discuss in more details in Sec.\ \ref{super}. Still, the identification of the oscillation frequency from the convolution \pref{E_omega} between the pump spectrum and the non-linear optical kernel is a robust feature, which holds regardless the specific frequency decay of the kernel away from resonance. In addition, the same description holds when instead of the changes of the transmitted field one measures the changes in the reflected one, as done e.g.\ in Refs \onlinecite{cavalleri_natphys16,cavalleri_science18}. Also in this case, the difference between the reflected probe beam with and without the pump originates from the non-linear current, which in turn depends on the non-linear kernel. As discussed below Eq.\ \pref{deltaE_tpp}, the dependence on the probe field completely disappears when the transmitted field is recorded as a function of the time delay $t_{pp}$ and the observation time $t_g$ is kept fixed. On the other hand, if one detects all the spectral component of the transmitted or reflected probe field at a fixed $t_{pp}$, this is equivalent to take the Fourier components of Eq.\ \pref{deltaE_tpp} with respect to $t_g$. In this case the oscillations of the differential signal as a function of $t_{pp}$ can only be evidenced by considering the spectrally-integrated pump-probe response, as shown indeed in Ref.\ \onlinecite{cavalleri_natphys16} in the case of plasmon excitations.

Finally, let us see how the present formalism explains also the phenomenon of high-harmonics generation from resonant excitations. As mentioned above, in this case one performs standard transmission experiments, using however a very intense pump field. From the point of view of the quartic-order action \pref{s4res}, this is equivalent to say that only components of the pump field appear, and since we are expanding the action only up to fourth order we are describing THG. If the pump is applied e.g.\ along the $x$ axis and one measures the component of the transmitted field in the same direction, it follows immediately that
\be
\lb{jnlthg}
J_x^{NL}(t)=-2e^4A_{pump}(t)\int\,dt' K_{xx;xx}(t-t')\left[A_{pump}(t')\right]^2.
\ee
When the current is collected along different crystallographic directions with respect to the pump field, one can obviously have mixed components of the non-linear kernel, that provide in general useful information on the nature of the collective resonance\cite{cea_prb16,shimano_prb17,cea_prb18}.  In all cases, it can be easily understood from Eq.\ \pref{Etr}  and \pref{jnlthg} that the Fourier spectrum of the transmitted field contains frequency components both at the pump frequency $\Omega_{pump}$, coming from the linear as well as from the non-linear current, and components around $3\Omega_{pump}$, due to the non-linear current \pref{jnlthg} only,
i.e.\ $ E^{THG}\propto J_x^{NL}\left(3\Omega_{pump}\right)$.
As discussed previously\cite{shimano_science14,aoki_prb15,cea_prb16,aoki_prb16,shimano_prb17,cea_prb18,shimano_cm19,silaev_cm19},  the THG intensity $I^{THG}$ scales once more with the non-linear kernel, and for a narrowband pulse one easily gets that 
$I^{THG}\sim | \int dt J_x^{NL}(t)e^{i3\Omega_{pump}t}|^2$ so that
\be
\lb{ithg}
I^{THG}(\Omega)\propto |K_{xx;xx}(2\Omega)|^2.
\ee
In this case, $\omega_{res}$ can be identified by varying $\Omega_{pump}$ until that the resonance condition $2\Omega_{pump}=\omega_{res}$ is fulfilled, as indeed expected for a sum-frequency process. 

By direct comparison between Eq.\ \pref{narrow} and Eq.\ \pref{ithg} one sees that there exists a general one-to-one correspondence between the spectral components of the differential transmitted field $\delta E_{probe}(t_{pp})$ and the spectral components of the transmitted one $E^{tr}$. Indeed, in full generality a peak in the transmitted component at $(n+1)\Omega_{pump}$, with $n$ integer,  correspond to oscillations at $n\Omega_{pump}$ in $\delta E_{probe}(t_{pp})$. They are both enhanced when a resonance $\omega_{res}$ exists in the non-linear optical kernel of order $n+1$, $K^{n+1}(\omega)$, at the matching condition $\omega_{res}=n\Omega_{pump}$. On this respect, the recent observation\cite{wang_natphot2019} in superconducting Nb$_3$Sn of odd harmonics in the oscillations of the differential field $\delta E_{probe}(t_{pp})$ is a remarkable result. Indeed, as we shall see in Sec.\ \ref{super}, in a SC the non-linear optical kernel contains only resonances at even multiples of the pump frequency, so that odd harmonics in $\delta E_{probe}(t_{pp})$ are in general not expected. In Ref.\ \onlinecite{wang_natphot2019} such anomalous high-harmonic generation has been interpreted in terms of a finite symmetry-breaking momentum of Cooper pairs induced by the asymmetry of the pump profile inside the sample and by the spatial inhomogeneity of the SC phase. How to incorporate these effects in the present description is an interesting question that will require future work. Finally, it is also worth observing that this quasi-equilibrium approach requires that the pump field is intense enough to generate non-linear effects, but not too large to affect the existence of a resonance in $K(\omega)$. In the case of phase transitions, as for superconducting or CDW systems, this implies that $K(\omega)$ can still be computed in the broken-symmetry state. However, when the pump strength increases the ground states is progressively weakened, as shown in Ref.\ \onlinecite{shimano_prl13} for SC NbN. At even larger peak electric fields, as in the case of SC Nb$_3$Sn investigated in Ref.\ \onlinecite{wang_natmat2018}, the pump pulse destroys the SC condensate and one eventually access a perturbed non-equilibrium state lasting for long time scales, but oscillations due to SC resonances disappear. 

The derivation done so far only relies on the existence of contribution to the non-linear response of the form of Eq.\ \pref{s4res}, where the non-linear optical kernel $K(\omega)$ has a well-defined resonance at $\omega_{res}$. In the next sections we will consider some specific examples where this condition is fulfilled, showing how the present scheme can be applied to several experimental results reported for various systems.
 
\section{Pump-probe spectroscopy of phonons}
\label{phonons}

As mentioned in the Introduction, it has been long ago recognized that coherent oscillations at the optical phonons frequency induced by short pulses in the eV range can be understood as a difference-frequency ISRS process\cite{merlin_SSC97}. More recently, it has been experimentally and theoretically discussed the possibility to induce phonon oscillations also by means of intense THz pulses in wide-band insulators, showing the difference between a DFP and a SFP\cite{kampfrath_prl17,maehrlein_prb18,johnson_prl19}. 
Here we want to recast these results within the picture presented in the previous Section, with the aim to further clarify how the time evolution of phonon displacement field can be correlated with the dependence of the transmitted field as a function of the pump-probe time delay and how the spectral features of the applied pulses influence the excitation process.

To fix the notation, we will focus on a generic band insulator, denoting with $c_{\bk,c}$, $c_{\bk,v}$ the annihilation operators of electrons in the conduction and valence band, respectively. On very general ground, we can assume that the phonon is coupled to density-like excitations in both bands, and that, for the sake of simplicity, the electron-phonon coupling $g$ does not depend on the band index and momentum:
\bea
H&=&\sum_\bk \e_c(\bk) c^\dagger_{c,\bk}c_{c,\bk}+\e_v(\bk) c^\dagger_{v,\bk}c_{v,\bk}+\nn\\
\lb{hel}
&+&\sum_\bq \omega_0 b^\dagger_\bq b_\bq+g\sum_{\bk,\bq,a,b} (b^\dagger_\bq+b_\bq) c^\dagger_{a,\bk+\bq}c_{b,\bk},
\eea
where $b_\bq$ denotes the phonon annihilation operator, and $a,b=v,c$ labels the valence and conduction band, respectively.  As usual\cite{mahan}, the gauge field couples to the electronic degrees of freedom via paramagnetic and diamagnetic terms, such that $H(\bA)\approx H(\bA=0)+\bA\cdot {\bf j}+\frac{1}{2}\sum_{\a\b}\rho_{\a\b}A_\a A_\b$, where the long-wavelength limit $\bq\simeq 0$ for the gauge field has been applied. In full generality, the current and diamagnetic operators can be expressed in the band basis as:
\bea
j_\a&=&\sum_{a,b=c,v} v_{ab}^\a c^\dagger_{a,\bk}c_{b,\bk}, \\
\rho_{\a\b}&=&\sum_{a,b=c,v} \tau^{\a\b}_{ab} c^\dagger_{a,\bk}c_{b,\bk},
\eea
where $\a,\b$ denote spatial indexes, and we consider the most general case where the velocity and diamagnetic terms have both intra-band and inter-band contributions. 

\begin{figure}
\centering
\includegraphics[scale=0.8]{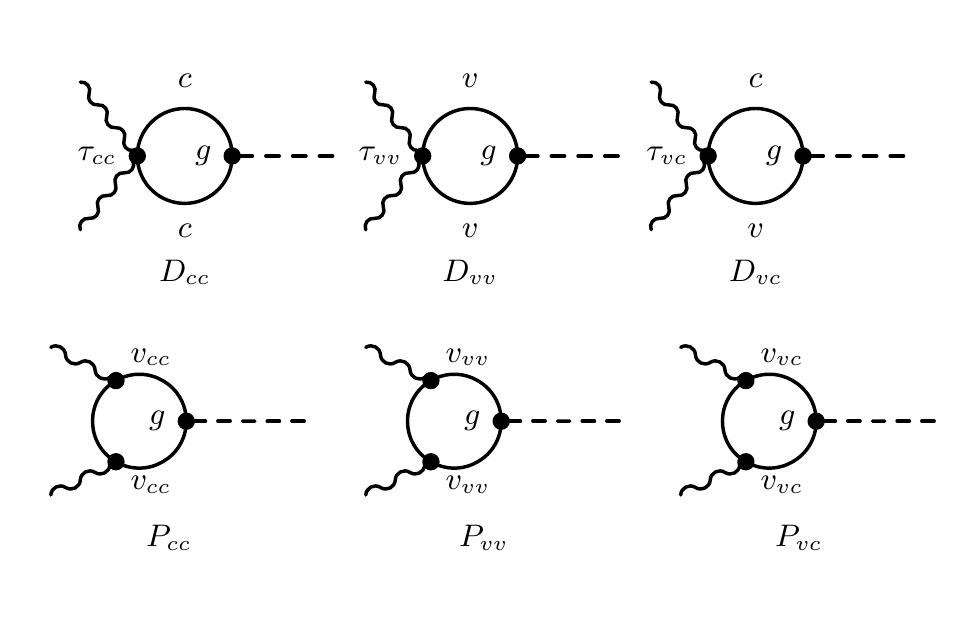}
\caption{
Fermionic bubbles connecting the gauge field (wavy lines) with the phononic field (dashed lines). Solid lines denote fermionic Green's function for the valence/conduction bands, as explicitly labeled in the diamagnetic diagrams on the top row. The vertexes identify the insertion either of a velocity $v_{ab}$ or a diamagnetic $\tau_{ab}$ term, which can both have intra-band and inter-band components, with $a,b=v,c$ denoting valence and conduction bands, respectively. }
\label{fig-diagrams}
\end{figure}

In order to derive the effective action for the gauge field, we resort to the functional-integral formalism. Since the model \pref{hel} is Gaussian in the fermionic variables, they can be strictly integrated out, leading to an effective action which depends on the phonon variables and on the gauge field only. By introducing as usual the displacement field $Q=(b+b^\dagger)$ for the phonon operators at $\bq=0$, as relevant for the long-wavelength excitations induced by a uniform e.m. field, the effective action then reads:
\bea
S&=&\sum_n \frac{1}{2}\left( \frac{\Omega_n^2+\omega_0^2}{2\omega_0}\right) |Q(i\Omega_n)|^2+\nn\\
&+&\sum_{n,m}e^2A_x(i\Omega_n)A_x(i\Omega_m)P(i\Omega_n,i\Omega_m) Q(i\Omega_n+i\Omega_m),\nn\\
\lb{saq}
\eea
where $i\Omega_n$ are bosonic Matsubara frequencies and $P(i\Omega_n,i\Omega_m)$ denotes the response function obtained by integrating out fermions. Even though the last term of Eq.\ \pref{saq} is quadratic in the gauge field, as we shall see below it leads to a fourth-order contribution to the electromagnetic action once the phonon field is integrated out, leading again to a structure similar to Eq.\ \pref{s4res}. Additional $\OO(A^4)$ terms mediated by the electrons have been discarded since the non-linear electronic kernel is not expected to have any resonance.
 
Even without an explicitly calculation, $P(i\Omega_n,i\Omega_m)$ can be conveniently expressed by means of the Feynman diagrams shown in Fig.\ \ref{fig-diagrams}. As explained above, the gauge field couples linearly to particle-hole excitations via the velocity vertices $v_{ab}$ and quadratically via the diamagnetic vertices $\tau_{ab}$. We can then identify diamagnetic contributions to $P$, as given by the first row of Fig.\ \ref{fig-diagrams}, and paramagnetic ones, in the second row. For a band insulator, it easily follows that the first two diamagnetic ($D_{vv}$ and $D_{cc}$) and paramagnetic ($P_{vv}$ and $P_{cc}$) contributions, corresponding to intra-band excitations within the valence/conduction band, vanish at $\bq=0$, since particle-hole excitations within a fully filled/empty band are not permitted. One is then left with the diamagnetic $D_{vc}$ and paramagnetic $P_{vc}$ diagrams, describing inter-band excitations between the valence and conduction band. To make a closer connection between the present scheme, where one computes the response with respect to $\bA$, and the ab-initio approach of Refs \onlinecite{kampfrath_prl17,maehrlein_prb18,johnson_prl19}, where the response with respect to the electric field $\bE$ is derived\cite{baroni_prb91}, we can rewrite the second term of Eq.\ \pref{saq} as 
\bea
\lb{seq}
-\sum_{n,m}e^2E_x(i\Omega_n)E_x(i\Omega_m)R(i\Omega_n,i\Omega_m) Q(i\Omega_n+i\Omega_m),\nn\\
\eea
where we  defined $R(i\Omega_n,i\Omega_m)=P(i\Omega_n,i\Omega_m)/\Omega_n\Omega_m$. The crucial observation now is that in a wide-band insulator the kernel $R$ appearing the Eq.\ \pref{seq} is weakly dependent on the external frequencies, as long as they are smaller than the band gap $\Delta$\cite{baroni_prb91,maehrlein_prb18}. Moreover, one can easily see that when this condition is fulfilled the function $R$ is real and it coincides with the Raman tensor of the phonon\cite{baroni_prb91} $R\equiv \pd \chi_{xx}/\pd Q$, where $\chi_{xx}$ is the electric susceptibility. In a small gap semiconductor or in a semimetal this approximation is instead not good and one can have a non-trivial dependence of the electron-phonon kernel $R$ on both frequencies. For example, within the context of graphene-based materials the frequency dependence of the electron-phonon infrared  kernel has been widely discussed within the context of the infrared response\cite{cappelluti_prb10,cappelluti_prb12,bistoni_2D19}, where it has been shown that the phonon effective charge can be greatly enhanced when the interband transitions macth the phonon frequency. Similar effects could be present for the Raman kernel, so for the moment we will consider only the case where the band gap is the largest energy scale in the problem. 

From the diagrammatic point if view, making the derivative with respect to the phonon field $Q$ is equivalent to the insertion of a phonon line in the electronic bubbles describing the response to $\bA^2$. This is what happens for all the diagrams in Fig.\ \pref{fig-diagrams}, that can be thus identified as the derivative of the current-current response function $\Pi_{xx}$ with respect to $Q$. By further using the connection\cite{mahan,baroni_prb91} between the electric susceptibility and the current-current response function, i.e.\ $\chi_{ij}(\omega)=-\Pi_{ij}(\omega)/\omega^2$, one understands the equivalence between the two approaches as far as the differences between the two incoming frequencies are negligible with respect to the bandgap. By then assuming that $R$ is real and by performing the analytical continuation to real frequencies, the effective action \pref{saq}-\pref{seq} reduces to:
\bea
S&=&\frac{1}{2}\int d\Omega \left( \frac{\Omega^2-\omega_0^2}{2\omega_0}\right) |Q(\Omega)|^2+\nn\\
&+&\int d\Omega d\Omega' e^2E_x(\Omega)E_x(\Omega-\Omega') R  \, Q(-\Omega)=\nn\\
\lb{sqereal}
&=&\int dt \,  \frac{\dot Q^2(t) -\omega_0^2 Q^2(t)}{4\omega_0}+e^2 R E_x^2(t) Q(t).
\eea
Once expressed in real time, Eq.\ \pref{sqereal} allows us to easily recover from the minimization condition $\delta S=0$  the standard equations of motion for the phonon displacement field\cite{merlin_SSC97,kampfrath_prl17,maehrlein_prb18,johnson_prl19}:
\be
\lb{eqtime}
\ddot Q(t)+\omega_0^2 Q(t)=2\omega_0 e^2 R E_x^2(t),
\ee
\\
where one directly sees that the force term in the rhs is quadratic in the electric field, as expected for a Raman-like excitation process. Eq.\ \pref{eqtime} can be easily solved by
\bea
Q(t)&=&e^2 R \int d\Omega e^{i\Omega t} D(\Omega) E^2_x(\Omega)=\nn\\
\lb{qt}
&=&e^2 R\int dt' D(t-t')E^2_x(t'),
\eea
where $E^2_x(\Omega)=\int d\omega E_x(\Omega)E_x(\Omega-\omega)$ and $D(\Omega)=2\omega_0/(\omega_0^2-(\Omega+i\delta)^2)$ is the phonon propagator, with $\delta$ an infinitesimal positive constant, that can be replaced by a finite broadening $\gamma$ to account for the finite phonon lifetime. 
 
 \begin{figure*}
\centering
\includegraphics[scale=0.5]{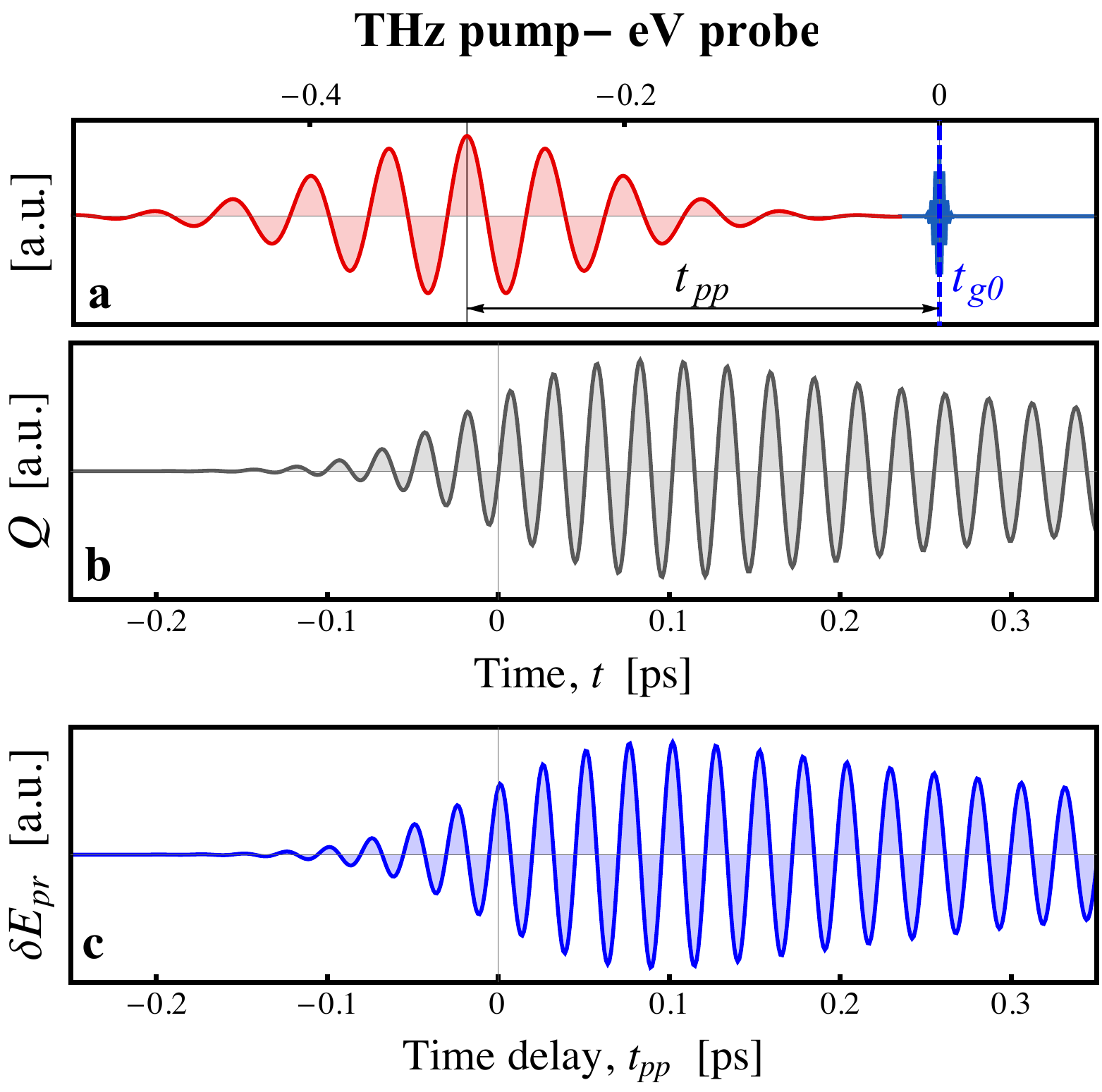}
\hspace{0.8cm}
\includegraphics[scale=0.5]{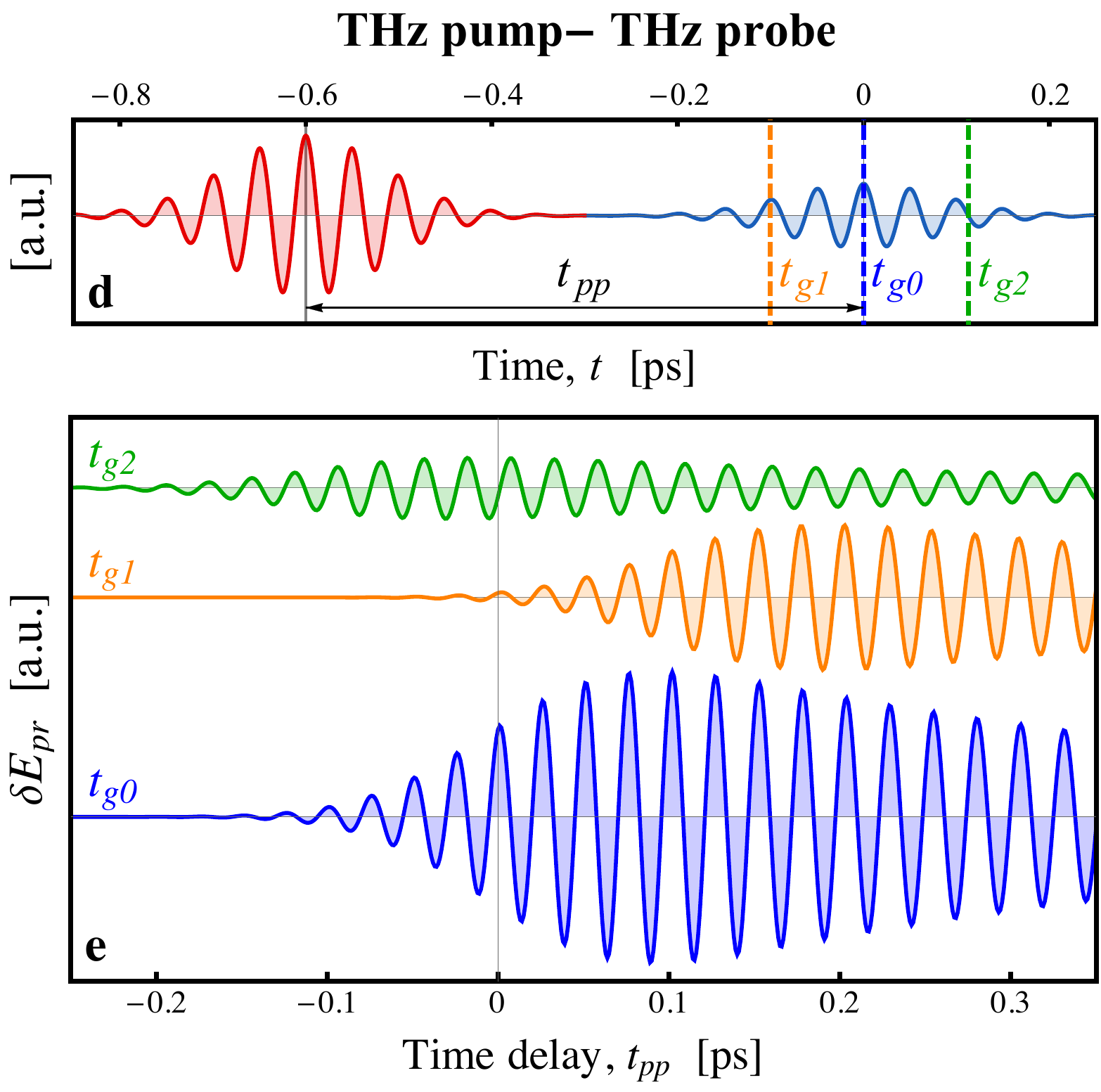}
\caption{Left: THz pump-eV probe detection of coherent phonon oscillations. Here the phonon mode oscillates at $\omega_0/(2\pi)=39.2$ THz, with spectral broadening $\gamma/(2\pi)=0.4$ THz. Panel \textbf{a} shows the time dependence of the pump and probe field (the observation time is set to $t_{g0}=0$). Since the duration of the probe field is much shorter than the period of the phonon, there is no appreciable difference between the time evolution of the phonon displacement $Q(t)$ induced by the pump, panel \textbf{b}, and the time dependence of the differential probe field as a function of the pump-probe time delay $t_{pp}$, panel \textbf{c}.  Right: THz pump-THz probe detection of coherent phonon oscillations. Panel \textbf{d} shows the time dependence of the pump and probe pulses. Panel \textbf{e} shows the differential transmitted probe field for three different values of $t_g$, as marked in panel \textbf{d}. Even though $\delta E_{probe}(t_{pp})$ always oscillates at the phonon frequency, the choice of $t_g$ affects the amplitude and phase of the signal.}
\label{fig-phonon} 
\end{figure*}

The solution of the equations of motion for $Q(t)$ is relevant to understand the excitation process induced by the pulse, and its behavior for insulating system perturbed via THz fields has been recently detailed in Refs \onlinecite{kampfrath_prl17,maehrlein_prb18,johnson_prl19}. However, computing explicitly $Q(t)$ does not provide yet the behavior of the differential probe field as a function of the pump-probe time delay $t_{pp}$, that is the actual measured quantity. In order to achieve this goal one should proceed further and derive the quartic-order action for the electric field only, in analogy with the derivation outlined in the previous Section. This can be easily done from Eq.\ \pref{sqereal} by integrating out the phonon field. For the sake of generality, we can keep the explicit dependence of the Raman tensor on the direction of the electric field, so that the last term in Eq.\ \pref{sqereal} reads $R_{\a\beta}E^2_{\a\beta}(t)$. In full analogy with Eq.\ \pref{s4res}, one then finds that
\bea
S^{(res)}(\bE)&=& -\int d\Omega E^2_{\alpha\beta} (\Omega) e^4 R_{\alpha\beta}R_{\g\d} D(\Omega) E^2_{\g\d}(-\Omega)=\nn\\
 &=&-\int dt dt' E^2_{\a\b}(t) e^4 R_{\alpha\beta}R_{\g\d} D(t-t') E^2_{\g\d}(t').\nn\\
\eea
The previous expression has exactly the same structure as Eq.\ \pref{s4res} above, except for the change of variables between the gauge field $\bA$ and the electric field $\bE$. In particular, it shows that the resonant non-linear kernel in the case of a phononic excitation is just the phonon propagator $D(t)$ itself, weighted by the Raman tensor. In the geometric configuration of Eq.\ \pref{Ein}, i.e.\ for a pump field along $y$ and a probe field along $x$, we can derive $\delta E_{probe}(t_{pp})$ as done in the previous Section, finding again that the differential probe field depends on the non-linear current $J_x^{NL}(t)$ only. Since $J_x^{NL}(t) = \pd P_x^{NL}/\pd t$, where $P_x^{NL}(t)=-\pd S^{(res)}/\pd E_x$ is the non-linear polarization, we easily get the equivalent of Eq.\ \pref{deltaE_tpp} for the phonon excitation:
\begin{widetext}
\be
\lb{deltaEQ}
\delta E_{probe}(t_g;t_{pp})\propto \frac{\pd E_{probe}}{\pd t}(t_g)\int dt' D(t_g+t_{pp}-t')\bar E^2_{pump}(t')+E_{probe}(t_g)\int dt' \dot D(t_g+t_{pp}-t')\bar E^2_{pump}(t'),
\ee
\end{widetext}
where the proportionality factor accounts for constant terms (including the Raman tensor) dropped out for the sake of compactness, and $E_{pump}(t)=\bar E_{pump}(t+t_{pp})$ has been rescaled to explicitly account for the pump-probe delay. Since phonon resonances are usually very sharp, i.e.\ $\gamma\ll \omega_0$, it simply follows that the Fourier transform of Eq.\ \pref{deltaEQ} with respect to $t_{pp}$ at fixed $t_g$ scales as
\be
\lb{deltaEOQ}
\delta E_{probe}(\omega)\propto R_{xx}R_{yy}D(\omega)\bar E^2_{pump}(\omega).
\ee
Eq.s\ \pref{deltaEQ}-\pref{deltaEOQ} are fully equivalent to Eq.\ \pref{deltaE_tpp}-\pref{E_omega}, showing that the phonon field can be excited only if it is Raman active {($R\ne0$)} and if there is a finite overlap between the spectral components of the squared pump field and the phonon frequency. Since the phonon response is peaked at $\Omega\simeq \omega_0$, in full analogy with the discussion below Eq.\ \pref{asquared}, one finds that for a broadband pump $\delta E_{probe}\propto \bar E^2_{pump}(\omega_0)$, so that $\delta E_{probe}(t_{pp})$ oscillates at the phonon frequency $\omega_0$. The underlying mechanism is again different for a pulse in the visible or in the THz frequency range, as recently discussed in Ref.\ \onlinecite{maehrlein_prb18}. In the former case, a finite component  of $\bar E^2(\Omega)$ around $\omega_0$ occurs because of the frequency mismatch between two incoming photons, and the phonon is excited via a DFP. In the latter case, one has to sum-up two THz photon frequencies to excite the phonon, thus dealing with a SFP. 

From the theoretical standpoint, the solution $Q(t)$ of the equation of motion \pref{qt} for the phonon field and the experimentally measured differential field $\delta E_{probe}(t_{pp})$ represent two distinct quantities. However, qualitative differences can be only appreciated if the probe field is in the THz. Indeed, both $Q(t)$ and $\delta E_{probe}(t_{pp})$ display oscillations at the phonon frequencies, as explicitly shown in Fig.\ \ref{fig-phonon}, where we solve Eq.\ \pref{qt} for $Q(t)$ and Eq.\ \pref{deltaEQ} for $\delta E_{probe}(t_{pp})$ using a proper set of parameters well reproducing the recent experiment in diamond of Ref.\ \onlinecite{kampfrath_prl17}. To make a closer connection with the theoretical analysis of Refs \onlinecite{kampfrath_prl17,maehrlein_prb18}, we model here the electric field by using an expression analogous to Eq.\ \pref{eqa} above, i.e.\ 
\be
\lb{epump}
\bar E_{pump}=E^0_{pump}e^{-\left ({2t\sqrt{\ln2}}/{\tau_{pump}}\right)^2}\cos(\Omega_{pump}t),
\ee
with $\Omega_{pump}/2\pi=20$ THz and $\tau_{pump}=0.2$ ps. For the VIS probe field we take instead $\Omega_{probe}/2\pi=$ 400 THz and $\tau_{probe}=7$ fs, as used in the experiment of Ref.\ \onlinecite{kampfrath_prl17}, while the Raman-active phonon is modeled as in Eq.\ \pref{ksharp}, with $\omega_0/2\pi=39.2$ THz and $\gamma=0.01\omega_0$.  For a probe  field in the visible, the pulse duration is much shorter than the time scale over which the phonon propagator varies in time, meaning that one can essentially set $t_g\simeq 0$ in Eq.\ \pref{deltaEQ}, see Fig.\ \ref{fig-phonon}a. In this situation, as depicted in Fig.\ \ref{fig-phonon}b-c no relevant difference can be appreciated between the time variation of $Q(t)$ and the pump-probe time-delay dependence of $\delta E_{probe}(t_{pp})$, apart for the overall amplitude of the two signals, scaling as $R$ and $R^2$ respectively. On the other hand, for a probe field in the THz range one can vary $t_g$ along the pulse duration (Fig.\ \ref{fig-phonon}d) and observe the changes in phase and amplitude of the oscillations in $\delta E_{probe}(t_{pp})$, as shown in Fig.\ \ref{fig-phonon}e where the probe field has the same spectrum and time evolution of the pump. It is worth noting that $\bar E_{pump}^2(\omega)$ for the THz pump used in Ref.\ \onlinecite{kampfrath_prl17} identifies a relatively narrow peak around $2\Omega_{pump}$, and since the phonon resonance is extremely sharp $\gamma_0/\omega_0\simeq 0.01$ one needs to go very close to the SFP resonance condition $2\Omega_{pump}\simeq \omega_0$ in order to excite the phonon, as indeed observed in Ref.\ \onlinecite{kampfrath_prl17}. On the contrary, in the case of Ref.\ \onlinecite{johnson_prl19} the squared pump spectrum $\bar E_{pump}^2(\omega)$ is much broader, and thus overlaps with several Raman-active phonons. In this case, according to Eq.\ \pref{deltaEOQ} the intensity of the oscillatory signal depends not only on the relative Raman cross-sections, but also on the value of $\bar E_{pump}^2(\omega_0^i)$ at the $\omega_0^i$ frequencies of the various phonon modes. More generally, if we model directly the pump electric field as in Eq.\ \pref{epump} we can strictly estimate from Eq.\ \pref{deltaEOQ} the relative efficiency of the process for a pulse in the IR-VIS or in the THz range. Indeed, in analogy with Eq.\ \pref{asquared},  we simply find that $E^2(\omega\approx 0)\simeq 2\tau_{pump} E_0^2$ and $E^2(\omega\approx 2\Omega_p)\simeq \tau_{pump} E_0^2$. Since for an eV pump what matters is the DFP, i.e.\ $E^2(\omega\approx 0)$, while for a THz pulse $E^2(\omega\approx 2\Omega_p)$ enters Eq.\ \pref{deltaEOQ} for the SFP, we deduce that for two pulses with the same value of the peak field $E_0$, one has
\be
 \frac{\delta E_{probe}^{THz-pump}}{\delta E_{probe}^{eV-pump}}\simeq \frac{\tau_{pump}^{THz}}{2\tau_{pump}^{eV}}.
 \ee
As a consequence a pump in the THz, with a typical duration of fractions of picoseconds, is about one or two orders of magnitude more efficient than a visible pulse, with a typical duration of tens of femtoseconds. 

It is also worth mentioning that a possible alternative mechanism to the one discussed here to induce coherent phonon oscillations is via the so-called ionic Raman scattering\cite{wallis_prb71,cavalleri_irs_natphys11,cavalleri_prb14,spaldin_prl17,johnson_prl19}. In this case the pump electric field excites two infra-red active optical phonons, that can in turn couple and transfer energy to a Raman-active one, thanks to the presence of anharmonic (trilinear) phononic interactions. From the point of view of the present formalism, this implies that Eq.\ \pref{deltaEOQ} above is replaced by a more complex convolution of the Raman-active phonon propagators times the squared pump field and the IR-active phonon propagator. As discussed in Refs \onlinecite{cavalleri_irs_natphys11,cavalleri_prb14,spaldin_prl17,johnson_prl19} the ionic Raman scattering and the stimulated Raman scattering described so far can be distinguished by their different sensitivity to the light polarization or by using a pump-probe protocol with two different THz pulses\cite{johnson_prl19}. 

Notice that in the recent measurements of Refs \onlinecite{kampfrath_prl17,johnson_prl19} the experimental set-up is slightly different than the orthogonal cross-polarization transmission scheme described here. However, the experimentally measured quantity is always proportional to the non-linear current and the above arguments can be easily generalized, since the general mechanism explaining the excitation and detection of the phonon remains unchanged. So far, we only considered the case of large-gap band insulator, where the Raman tensor at the pump frequencies is always real, both for light pulses in the visible and in the THz. However, the situation can be rather different for small-gap insulators or normal metals, since in this case a finite density of particle-hole excitations is present at the pump frequencies, and the Raman tensor may acquires an imaginary part. In this case, for VIS-IR pulses one expects a displacive response\cite{merlin_SSC97}, and the frequency dependence of the Raman tensor itself can induce significant differences between the pump-probe detection of phonons via a DFP or a SFP. In particular, for THz pulses the intra-band diamagnetic $D_{cc}$ and paramagnetic $P_{cc}$ processes depicted in Fig.\ \ref{fig-diagrams} become relevant, and they must be explicitly computed for the specific band structure of the system under investigation.  As we shall see in the next Section, the ongoing theoretical debate on the nature of the SC excitations detected via the pump-probe protocols relies indeed in the analysis of low-energy intra-band processes triggered by THz pulses in metallic systems which undergo a SC transition.

\section{Pump-probe spectroscopy of superconducting modes}
\label{super}

As already mentioned, recent experiments in SC systems have shown that in the cross-polarized configuration of Eq.\ \pref{Ein} the differential field $\delta E_{probe}(t_{pp})$ shows marked oscillations when entering in the SC state\cite{shimano_prl12,shimano_prl13,shimano_science14,giorgianni_natphys19,wang_natphot2019}. At the same time, below the SC critical temperature $T_c$ a marked THG has been observed\cite{shimano_science14,cavalleri_science18,keiser_cm19}, which has been also ascribed to SC collective modes\cite{shimano_science14,aoki_prb15,cea_prb16,aoki_prb16,cea_leggett_prb16,shimano_prb17,cea_prb18,shimano_cm19,silaev_cm19,wang_natphot2019}. For the case of conventional superconductors, like thin films of NbN, both pump-probe protocols and THG measurements show that the relevant frequency is twice the value of the SC gap $\Delta$. As explained in Sec.\ \ref{general-theory}, the THG is potentially easier to be understood theoretically, since this is an equilibrium transmission measurement whose intensity scales with the non-linear optical kernel, see Eq.\ \pref{ithg}. Indeed, the ongoing debate on the literature focuses not on the relation \pref{ithg}, but rather on the identification of the collective SC excitations which give the largest contribute to $K(\omega)$.  In particular, using again a diagrammatic description, one has two main contributions to $K=\chi^{DF}+\chi^{H}$ in the $S^{(4)}$ action of Eq.\ \pref{s4res}:  the contribution $\chi^{DF}$ of lattice-modulated density fluctuations (LMDF), which probes  Cooper pairs excitations, and 
the contribution $\chi^{H}$ of the SC Higgs (amplitude) mode, arising from the RPA vertex correction in the pairing channel. With respect to the discussion of the previous Section, we are here focusing only on intra-band processes occurring in the bands at the Fermi level. 

The calculations of $K_{ij}=\chi^{DF}_{ij}$ has been detailed in several previous works\cite{cea_prb16,shimano_prb17,cea_prb18,silaev_cm19,shimano_cm19} and we will just report here the result in the single-band case: 
\be\lb{chiCP}
\chi_{ij}^{DF}(\o)=\frac{\Delta^2}{N_s}\sum_\mathbf{k}
\frac{\partial^2_{k_i}\varepsilon_\bk\partial^2_{k_j}\varepsilon_\bk}{E_\bk \left[(\o+i\g)^2-4E_\bk^2 \right]}\tanh(E_\bk/2T),
\ee 
where $N_s$ is the number of lattice site, $\Delta$ the SC order parameter,  $\xi_\bk=\varepsilon_\bk-\mu$ with $\varepsilon_\bk$ electronic band dispersion and $\mu$ chemical potential,  $E_\bk=\sqrt{\xi_\bk^2+\Delta^2}$ is the quasiparticle dispersion in the SC state and $\g$ a finite broadening. Apart for the tensorial nature implicit in the polarization factors $\partial^2_{k_i}\varepsilon_\bk$, $\chi^{DF}_{ij}$ is substantially equivalent to the BCS density-density correlation function $F(\omega)$, that at $T=0$ can be estimated as:
\bea
\lb{fomega}
F(\omega)&=&\frac{\Delta^2}{N_s}\sum_\mathbf{k}
\frac{1}{E_\bk \left[(\o+i\g)^2-4E_\bk^2 \right]}\simeq\nn\\
&\simeq&
{\Delta^2N_F} \int_{-\o_D}^{\o_D}\frac{\,d\xi/\sqrt{\xi^2+\Delta^2}}{\left[(\o+i\g)^2-4(\xi^2+\Delta^2)\right]}=\nn\\
&=& -\frac{2\Delta^2N_F}{m^2}\frac{\tan^{-1}\left[\frac{\o+i\g}{\sqrt{4\Delta^2-(\o+i\g)^2}}\frac{\o_D}{\sqrt{\o_D^2+\Delta^2}}\right]}{(\o+i\g)\sqrt{4\Delta^2-(\o+i\g)^2}}, \nn\\
\eea
where $\o_D$ gives an upper cut-off for the pairing, like e.g.\ the Debye frequency for phonon-mediated superconductivity. 
For the Higgs contribution we have that $\chi^H(\omega)= \chi^2_{A^2\Delta} \delta\Delta^2(\omega)$, with $\delta \Delta^2(\omega)$ the propagator for the Higgs mode and $\chi_{A^2\Delta}$ the Raman-like coupling between the gauge field and amplitude fluctuations $\delta \D$, which explicitly reads\cite{cea_prb16,shimano_prb17,cea_prb18,silaev_cm19,shimano_cm19}:
\be
\lb{higgs}
\chi^H(\omega)=\chi^2_{A^2\Delta} \delta\Delta^2(\omega)=\frac{\chi^2_{A^2\Delta}}{[(\omega+i\delta)^2-4\Delta^2]F(\omega)},
\ee
where the simple single-band case on a square lattice is considered, so that $\chi_{A^2\Delta}$ does not depend on the light polarization. The real and imaginary parts of the function $K(\omega)$ in the weak coupling limit are shown in the Appendix \ref{longtime} (see Fig.\ \ref{kernel}). In the case of zero broadening we recover a square-root singularity of the real part of $K$ at $\omega=2\Delta$, while the imaginary part remains zero for $\omega<2\Delta$ and diverges as $\sim(\omega-2\Delta)^{-1/2}$ for $\omega\to2\Delta^+$. The appearance of an imaginary part only at $\omega>2\Delta$ at $T=0$ reflects the gapped nature of the electronic spectrum $E_\bk$ in a superconductor, where absorption is only possible when the incoming frequency overcomes the $2\Delta$ threshold for the particle-hole continuum. Since the Higgs fluctuations identify a  collective electronic mode, the function $F(\omega)$ appears also in the Higgs contribution \pref{higgs}, so both the LMDF and the Higgs show a square-root singularity at $2\Delta$, and not a simple Lorentzian-like resonance as in the case of phonons discussed in Sec.\ \ref{phonons}. While this has some implications for the long-time decay of the oscillations, it does not alter the general description of the pump-probe experiment described in Sec.\ \ref{general-theory} for what concerns the difference between broadband and narrowband pulses. 

In the metallic state $K(\omega)=0$, since it scales with $\bq=0$ charge fluctuations that vanish at finite frequency  due to charge conservation\cite{deveraux_review,cea_leggett_prb16,chubukov_raman_prb17}. As shown above, in the SC state the density-like response $F(\omega)$ computed at BCS level is not zero, since density is not conserved in the BCS approximation. When the Raman density scales as the full density, as it occurs for parabolic-like bands and symmetric Raman channels,  the gauge invariance is restored by properly adding to Eq.\ \pref{fomega} the contribution of phase and density modes at RPA level\cite{cea_leggett_prb16,chubukov_raman_prb17}. However, for a lattice system the lattice-modulated density fluctuations \pref{chiCP} do not follow the same conservation law, and a finite response is possible in the SC state even in the absence of disorder\cite{cea_prb16,chubukov_raman_prb17,shimano_prb17}.  Since both $\chi^{DF}$ and $\chi^H$ diverge at $2\Delta$ the two contributions cannot be distinguished by their resonance frequency. However, for a conventional BCS superconductor the Higgs-mode contribution is largely subleading with respect to density-like fluctuations\cite{cea_prb16,cea_prb18}, because $\chi_{A^2\Delta}$ is extremely small. In other words, even if the gauge field is able to excite the Higgs mode as soon as $\chi_{A^2\Delta}$ is non zero, the contribution of the Higgs fluctuations to the non-linear kernel $K(\omega)$, and then to the physically accessible quantity $\delta E_{probe}(t_{pp})$, is negligible. The situation can change, and eventually even be reversed, when strong interactions\cite{werner_prb16,aoki_prb16} are considered, even if this also implies a significant smearing of the resonance itself, challenging the understanding of the experiments. Recently, it has been also suggested that disorder can play a crucial role, by activating paramagnetic-like processes which are absent without disorder\cite{shimano_cm19,silaev_cm19,shimano_review19}. It is also worth noting that the two contributions differ from their dependence on the polarization of the pump light\cite{cea_prb16}, but the exact comparison with the experiments require an accurate modelling not only of the band structure\cite{shimano_prb17} but also of the pairing interactions\cite{cea_prb18}. Finally, when disorder is present all the terms in the action \pref{s4gen}  whose kernel depends on multiple frequencies are in general not zero. In particular, it has been shown in \cite{silaev_cm19} that the paramagnetic-like response shows an additional resonance below $2\Delta$. In general, while in the clean case the single-frequency approximation encoded in Eq.\ \pref{s4gen} is exact below $T_c$\cite{cea_prb16}, this is not necessarely true in the presence of disorder.  On the other hand, the analysis of Ref.\ \cite{shimano_cm19,silaev_cm19} suggests that the main effect of disorder is to enhance the $\chi_{A^2\Delta}$ prefactor in Eq.\ \pref{higgs}. In this case, the largest contribution to the kernel still retains the single-frequency form of Eq.\ \pref{higgs}, since electronic excitations just mediate the coupling of light to the Higgs mode, in full analogy with the case of the phonon response discussed in Sec. II. 

\begin{figure}
\centering
{\includegraphics[scale=0.5]{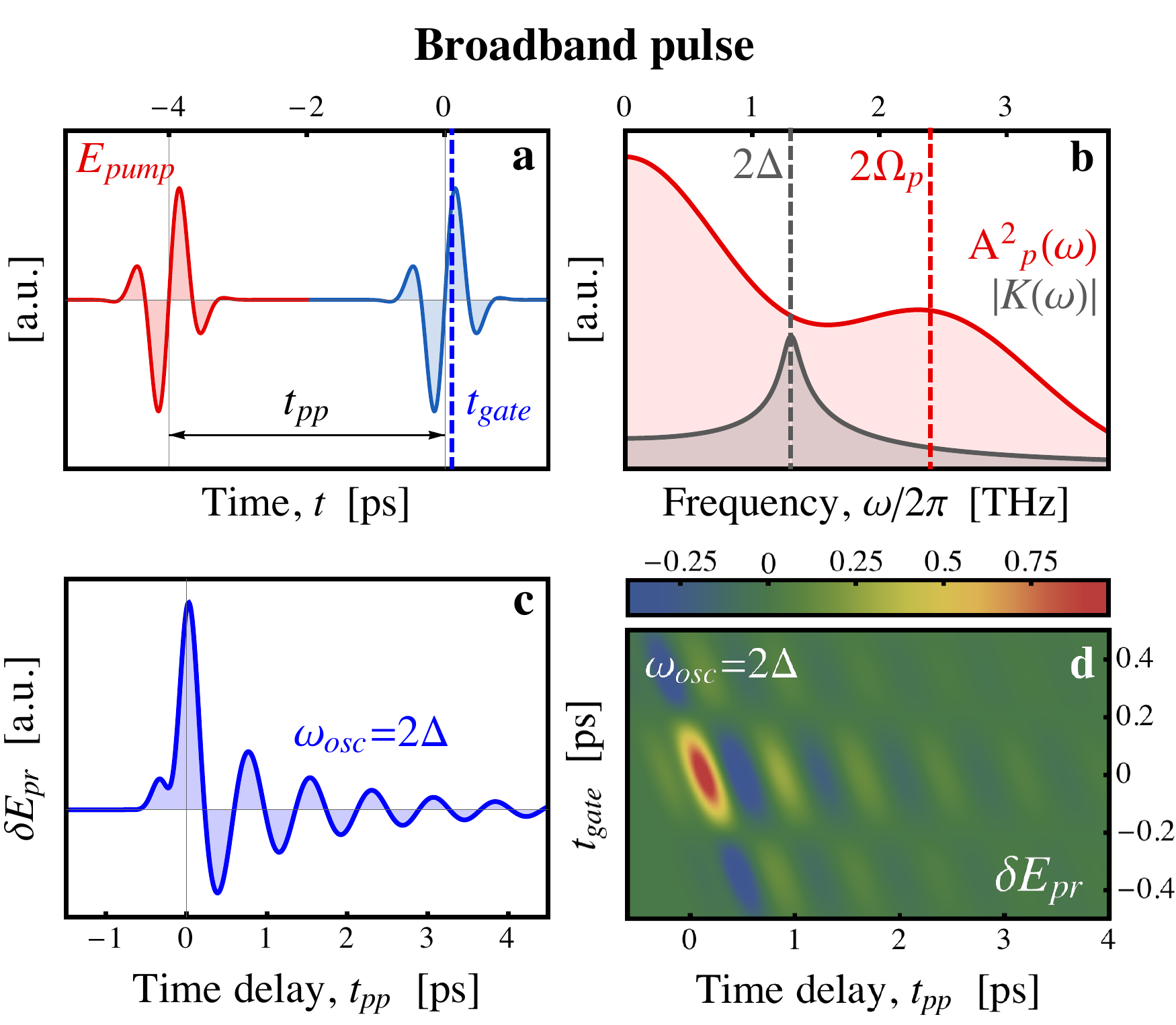}}
\caption{{\bf a.} Pump and probe fields used to simulate the experiments of Ref.\ \onlinecite{shimano_prl13}. {\bf b.} Power spectrum of the squared pump field $A^2_p(\omega)$ compared to the non-linear optical kernel $|K(\omega)|$ given by Eq.\ \pref{KconstDOS} for $\Delta=0.65$ THz. {\bf c.} Differential transmitted probe field as a function of the pump-probe delay $t_{pp}$ obtained by fixing the observation time at $t_{gate}=0.1$ ps. {\bf d.} $\delta E_{probe}(t_g;t_{pp})$ as a function of both the observation time $t_g$ and the time delay $t_{pp}$. 
}\label{experiment_broad_pump}
\end{figure}
In the following we will focus for simplicity on the case of a single-band BCS superconductor, where the DF contribution $\chi^{DF}$ is certainly the dominant one at weak disorder, with the aim to establish the connection between the non-linear kernel and the pump-probe response.  In particular, we will show that the experimental observations can be fully captured by using the same formalism of Sec.\ \ref{general-theory}, showing that the $2\Delta$ oscillations can also be explained using LMDF,  whenever these dominate the non-linear kernel. Since the band derivatives scales with the inverse electron mass $\left(\partial_i^2\varepsilon_\bk\right)\sim 1/m$, we will then approximate the non-linear optical kernel as 
\be
\lb{KconstDOS}
K(\o)\sim \frac{1}{m^2}F(\omega).
\ee
Let us start from the case of a broadband pump. Here both the pump and probe fields are modeled as in Eq.\ \pref{eqa},   with a central frequency of 1.2 THz and a duration of 0.64 ps, giving rise to the typical mono-cyclic time profile shown in Fig.\ \ref{experiment_broad_pump}a. The corresponding power spectrum of the pump squared  is shown in Fig.\ \ref{experiment_broad_pump}b along with $|K(\omega)|$ given by Eq.\ \pref{KconstDOS} for a gap value  $\Delta=0.65$ THz, consistent with the experiments performed on NbN SC films\cite{shimano_prl13,shimano_science14}. Since the power spectrum of $\bar A^2_{pump}(\omega)$ is much broader than the SC resonance, on the base of Eq.\ \pref{E_omega} $\delta E_{probe}(\omega)$ is peaked at $\omega\simeq \omega_{res}=2\Delta$, and the differential field $\delta E_{probe}(t_{pp})$ computed according to Eq.\ \pref{deltaE_probe} displays marked oscillations with a frequency $2\Delta$, see Fig.\ \ref{experiment_broad_pump}c, as indeed reported in Ref.\ \onlinecite{shimano_prl13} for a broadband pump pulse. 
As already mentioned in Sec.\ \ref{phonons}, and shown in Fig.\ \ref{fig-phonon}e for the case of the phonon resonance, the intensity and phase of the oscillations can be modulated by changing the observation time $t_g$. 
This effect is demonstrated in Fig.\ \ref{experiment_broad_pump}d, where we show in a contour plot the profile of $\delta E_{probe}$ as a function of both $t_{pp}$ and $t_{gate}$. Here one can clearly see that for different choices of $t_{gate}$ the oscillations with respect to the time delay $t_{pp}$ preserve their frequency at $2\Delta$, but they get progressively dephased, that is why the colored stripes of the plot assume a diagonal shape. A similar effect has been recently seen in broadband measurements of the SC Leggett mode\cite{giorgianni_natphys19}, and it would be interesting to test it experimentally also for the $2\Delta$ mode. 

\begin{figure}
\centering
\hspace{-0.4cm}
\includegraphics[scale=0.5]{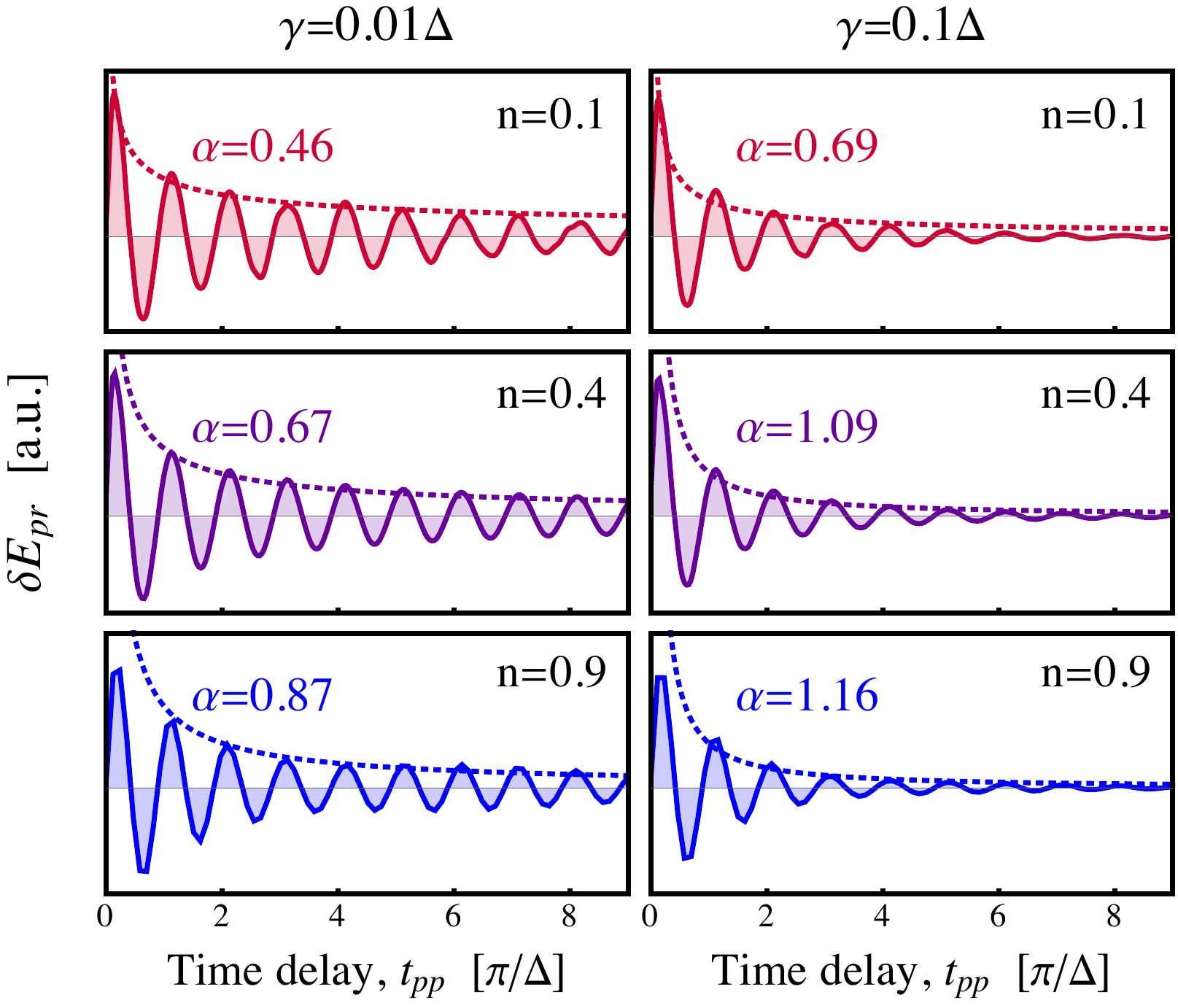}
\caption{
Time profile of the transmitted probe field in the SC case obtained within the attractive Hubbard model on the squared lattice at increasing filling $n=0.1, 0.4, 0.9$, and different values of the electronic broadening ($\gamma=0.01\Delta$, left column, $\gamma=0.1\Delta$, right column).
}\label{deltaE_Hubbard}
\end{figure}

While the frequency of the SC oscillations is well captured by the parabolic-band approximation \pref{KconstDOS} for the non-linear kernel, their long-time decay depends crucially on the band-structure details.  This can be again understood by relying on the general equation \pref{deltaE_probe} for the transmitted probe field. In the case of a broadband pulse, and setting for simplicity $t_g=0$,  we can safely approximate the long-time behavior of $\delta E_{probe}(t_{pp})$ with the time dependence on the non-linear kernel $K(t)$: 
\be\lb{deltaEsimK}
\delta E_{probe}(t_{pp})\sim K(t_{pp}).
\ee
For the ideal system with parabolic band structure, where $K(\omega)$ is given by the Eq.\ \pref{KconstDOS}, its Fourier transform at zero broadening $\gamma=0$ decreases as $K(t)\sim 1/\sqrt{t}$. This can be easily shown using the same argument usually invoked to explain the long-time decay of the Higgs mode\cite{volkov73,barankov_prl2006,dzero_prl2006,boris_prl2006}, as discussed in the Appendix \ref{longtime}. However, as soon as one considers a more realistic band structure the long-time decay of the non-linear optical kernel, and then of the differential probe field, is found to follow a power-law decay with a non-universal exponent
\be\lb{fit_function}
\delta E_{probe}(t_{pp})=\frac{A\cos(2\Delta t_{pp}+\phi)}{t_{pp}^\alpha}.
\ee 
We tested Eq.\ \pref{fit_function} by computing $\delta E_{probe}(t_{pp})$ for a prototype  tight-binding model on the square lattice, as the one discussed  in Ref.\ \onlinecite{cea_prb16}. We found that for a particle density near half-filling, where the effects of the density-of-states variations are stronger due to the proximity to a Van Hove singularity, the profile of the differential probe field follows closely Eq.\ \pref{fit_function}, with an exponent $\alpha\gtrsim1$, see Fig.\ \ref{deltaE_Hubbard}. It is worth noting that a power-law behavior of the long-time signal, with an exponent $\alpha$ ranging from $1$ to $3$, has been reported in the experiments on NbN films in Ref.\ \onlinecite{shimano_prl13}. 
Since a realistic band structure for NbN is near to half-filling\cite{shimano_prb17,cea_prb18}, the results of Fig.\ \ref{deltaE_Hubbard} indicate that deviations from the ideal $1/\sqrt{t}$ decay  could be captured by considering realistic band structures. 

\begin{figure}
\centering
\includegraphics[scale=0.5]{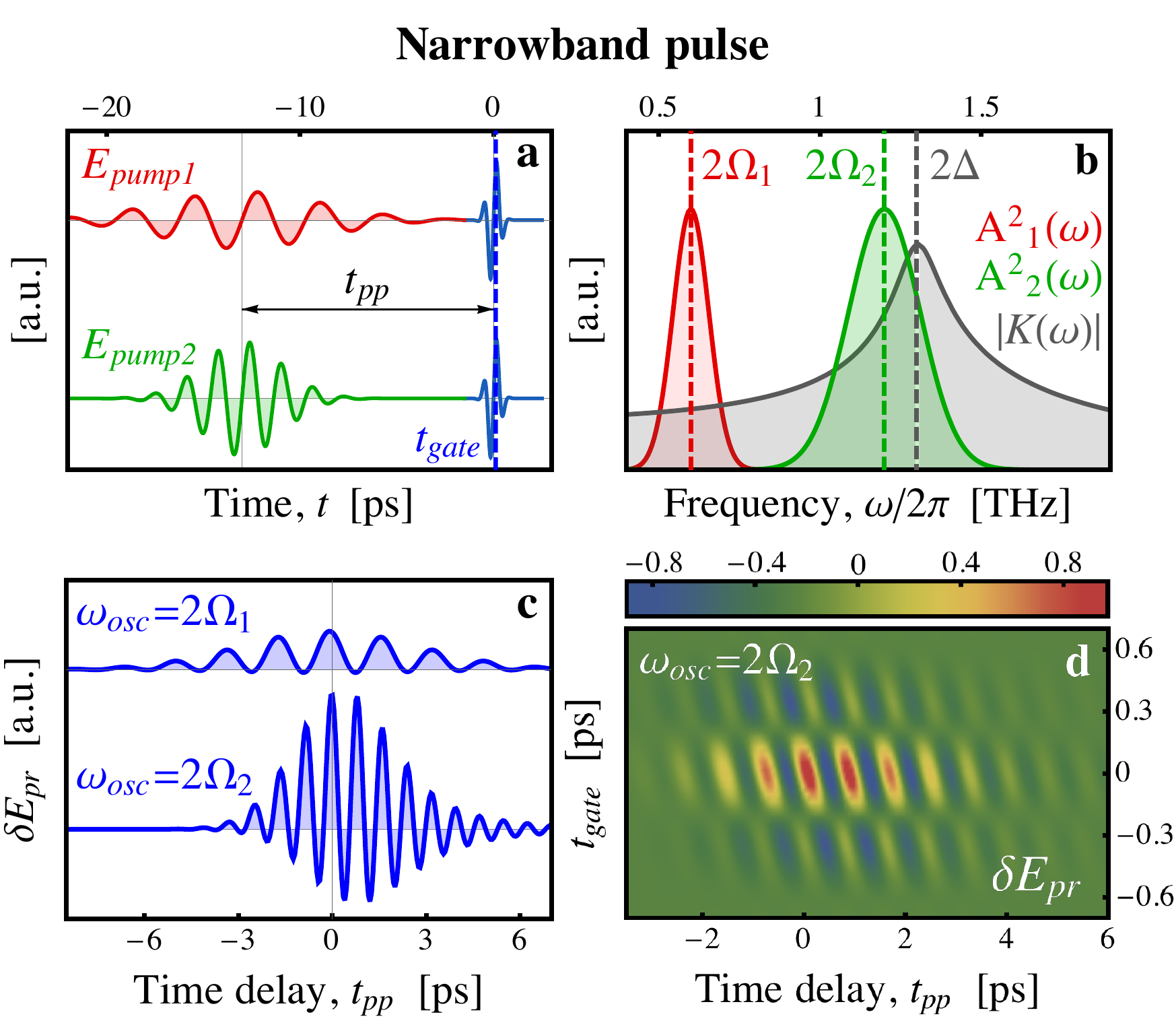}
\caption{{\bf a.} Pump fields (red and green lines) used to simulate the experiments of Ref.\ \onlinecite{shimano_science14} for two different central frequencies $\Omega_i$ and $\tau_i$. The probe field (blue line) is the same of Fig.\ \ref{experiment_broad_pump}a. {\bf b.} Power spectrum of the squared pump fields $A^2_{i}(\omega)$,  compared to the non-linear optical kernel $|K(\omega)|$ given by Eq.\ \pref{KconstDOS} for $\Delta=0.65$ THz. {\bf c.} Differential transmitted probe field as a function of the pump-probe delay $t_{pp}$ obtained by fixing the observation time at $t_{gate}=0.1$ ps. For both choices of the pump the oscillations are recovered at twice the pump frequency, with an enhanced amplitude when the resonance condition $2\Omega_2\approx 2\Delta$ is satisfied. {\bf d.} $\delta E_{probe}(t_g;t_{pp})$ as a function of both the observation time $t_g$ and the time delay $t_{pp}$. 
}\label{experiment_narrow_pump}
\end{figure}

Let us finally turn to the case of a narrow-band pump pulse. We model the pump with Eq.\ \pref{eqa} by setting $\Omega_{pump}/2\pi=0.3,0.6$ THz and $\tau_{pump}=9.5,4.8$ ps respectively, see Fig.\ \ref{experiment_narrow_pump}a, that reproduce the experimental configuration of Ref. \onlinecite{shimano_science14} with great accuracy.  In this case, the power spectrum of the squared pump field is considerably narrower than the non-linear kernel, see Fig.\ \ref{experiment_narrow_pump}b, so that according to Eq.\ \pref{deltaE_probe} $\delta E_{probe}(\omega)$ is dominated by twice the frequency of the pump field, weighted with the value of $K(2\Omega_{pump})$. As a consequence, as expected from Eq.\ \pref{narrow}, the differential field oscillates at $2\Omega_{pump}$, with a stronger amplitude when $\Omega_{pump}=0.6$ THz, see Fig.\ \ref{experiment_narrow_pump}c, such that the resonance condition with the maximum of the non-linear kernel is satisfied, i.e.\ $2\Omega_{pump}\simeq 2\Delta$. Also in this case our results not only capture the frequency of the oscillations of $\delta E_{probe}$ obtained experimentally in Ref.\ \onlinecite{shimano_science14}, but also their general profile as a function of $t_{pp}$, whose shape and time duration look very similar to that shown in the Ref. \onlinecite{shimano_science14}. In this case oscillations are much more long-lived, since they have essentially the duration of the pump pulse, and one can better resolve the phase and amplitude variations as a function of the observation time $t_g$, as shown in Fig.\ \ref{experiment_narrow_pump}d. 

As discussed at the beginning of this Section, there is still some ongoing theoretical debate in the literature on the nature of the collective SC mode responsible for the $2\Delta$ resonance in the non-linear optical kernel of a conventional superconductor. While the discussion has been focused so far on the THG\cite{shimano_review19}, the detailed derivation we gave above of $\delta E_{probe}(t_{pp})$ shows that the same problem exists for pump-probe experiments. Indeed, despite the fact that the experimental findings have been attributed so far uniquely to the oscillations of the Higgs mode\cite{shimano_prl13,shimano_science14}, we have shown that all the experimental features can be well reproduced by considering the effect of density-like fluctuations as well, once that these dominate the non-linear optical kernel. On the other hand, since the spectral features of the non-linear optical response \pref{higgs} due to Higgs fluctuations are qualitatively very similar to the ones of density fluctuations \pref{chiCP}, once the Higgs dominates the non-linear optical kernel the present approach explains equally well the difference between experiments done with broadband or narrowband pulses.  Unfortunately, for an ordinary BCS superconductor both the Higgs mode and charge fluctuations occur at the same frequency $2\Delta$, so the only way to discriminate them is to rely on an explicit calculation allowing one to decide which contribution dominate  $K(\omega)$. As we shall see in the next Section, the situation is instead different for the Higgs mode of a CDW system, since in this case charge and amplitude fluctuations occur at different frequency scales providing a tool to access them separately.

\section{Higgs mode of the charge-density-wave}

\begin{figure}
\centering
\hspace{-0.4cm}
\includegraphics[scale=0.5]{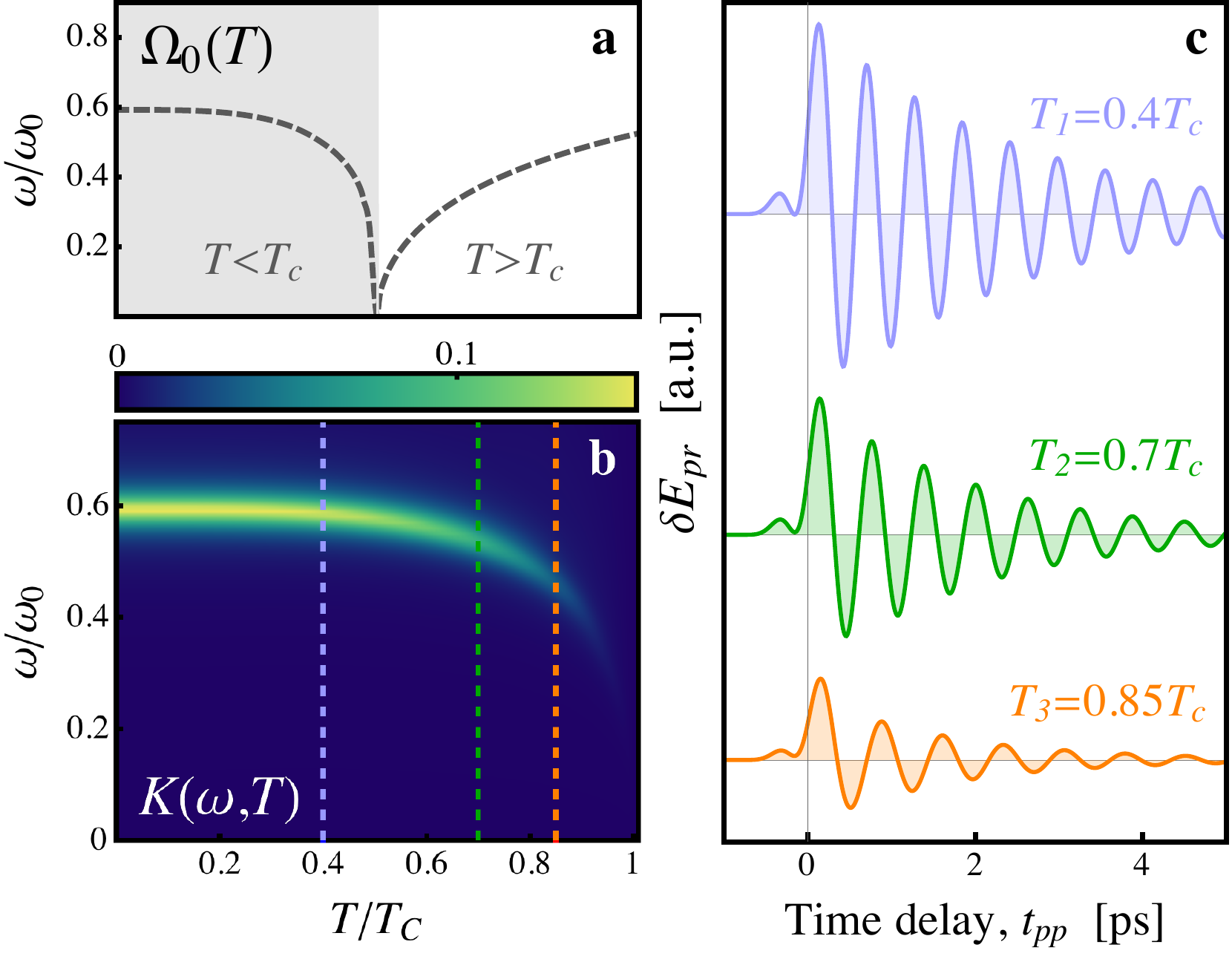}
\caption{{\bf a.} Temperature dependence of the renormalized phonon frequency $\Omega_0$ in units of the bare one $\omega_0$, as given by Eq.\ \pref{omegaup} at $T>T_{CDW}$ and Eq.\ \pref{omegaren} at $T<T_{CDW}$. {\bf b.} Raman kernel \pref{chi} of the amplitudon below $T_{CDW}$. Its value at the three selected temperatures shown by dashed lines its used to compute $\delta E_{probe}(t_{pp})$ according to Eq.\ \pref{E_omega}. {\bf c.} Differential field as a function of the pump-probe delay $t_{pp}$  for a broadband pulse with central frequency of $0.9$ THz and a duration of $0.8$ ps.}
\label{fig-cdw}
\end{figure}

In full analogy with the SC case, when a system undergoes a CDW transition a new electronic order parameter appears, whose ground-state value $\Delta_{CDW}$ is connected to the CDW gap in the electronic spectrum, and whose amplitude fluctuations define a massive collective mode. Within the conventional Peierls-like description of the CDW transition\cite{gruner_review,rice_ssc73}, the ordering is driven by an instability of the electronic charge susceptibility at the nesting wavevector $\bQ$, which causes the softening of the coupled phonon mode at $\bQ$, whose frequency $\Omega_0$ is renormalized by charge fluctuations and thus softens when $T$ approaches the critical temperature $T_{CDW}$.  Below $T_{CDW}$ the phonon couples to the amplitude (Higgs) fluctuations of the CDW order parameter\cite{rice_ssc74,brouwne_prb83,cea_cdw_prb14,grasset_prb18}, and its frequency $\Omega_0$ progressively increases by reducing the temperature, reaching a finite value at $T=0$ that is usually much smaller than the bare one $\omega_0$, see Fig.\ \ref{fig-cdw}a. This peculiar soft phonon mode, named "amplitudon" for its coupling to the CDW amplitude fluctuations, becomes Raman active below $T_{CDW}$\cite{klein_raman82,cea_cdw_prb14}, thus allowing for its detection with conventional Raman spectroscopy in several CDW systems, ranging from dichalcogenides\cite{tsang_prl76,chiang_prl03,measson_prb14,grasset_prb18,grasset_prl19} to tritellurides\cite{degiorgi_pnas13}. A microscopic derivation of the Raman kernel for the amplitudon within a simplified electron-phonon model has ben recently provided in Refs \onlinecite{cea_cdw_prb14,grasset_prb18}. The model system is based on the electron-phonon Hamiltonian:
\be
\lb{hmodel}
H=\sum_{\bk\sigma} \xi_{\bk}c^\dagger_{\bk\s}c_{\bk\s}+g\sum_{\bk\s}\gamma_{\bk}c^\dagger_{\bk+\bQ \s}c_{\bk\s}(b^\dagger_\bQ+b_{-\bQ}),
\ee
where $\xi_\bk=-2t(\cos k_x+\cos k_y)-\mu$ is the tight-binding dispersion on the square lattice, with $t$ the hopping parameter and $\mu$ the chemical potential, such that at half filling ($\mu=0$) the band dispersion displays perfect nesting at $\bQ=(\pi,\pi)$. As a consequence, the finite coupling $g$ to a phonon mode at $\bQ$ induces a CDW instability. Here $\gamma_\bk$ accounts for the possible momentum modulation of the CDW gap, and we use as an example a $d$-wave form factor $\gamma_\bk=(\cos  k_x-\cos k_y)/2$. As discussed in Refs \onlinecite{cea_cdw_prb14,grasset_prb18}, the present toy model allows one to capture the main features of several experimental observations, and it has been successfully applied to predict the pressure dependence of the amplitudon in Raman spectra of NbSe$_2$\cite{grasset_prb18}. The mean-field CDW order parameter $\Delta_{CDW}$ can be readily obtained from Eq.\ \pref{hmodel} as a solution of the self-consistent equation:
\be
\lb{selfc}
\Delta_{CDW}=\frac{2g^2\Delta_{CDW}}{\omega_0N}\sum_\bk \frac {\gamma_\bk^2}{E_\bk}\tanh(E_\bk/2T),
\ee
where $E_\bk=\sqrt{\xi_\bk^2+(\Delta_{CDW}\gamma_\bk)^2}$ is the quasiparticle dispersion below $T_{CDW}$.   As one can see, from the point of view of single-particle excitations the CDW transition has the same effect of the SC one, with the opening of a gap in the quasiparticle spectrum, so that the density-like fluctuations become resonant at $2\Delta_{CDW}$. However, since at the CDW transition a reconstruction of the lattice occurs, this reflects in the phononic spectrum. For both dichalcogenides\cite{grasset_prl19} and tritellurides\cite{degiorgi_pnas13} the CDW gap is expected to be significantly larger than the soft CDW phonon, so the non-linear kernel $K(\omega)$ in the THz range will be dominated in this case by the amplitudon, allowing one to detect Higgs fluctuation of the CDW order parameter at a scale $\Omega_0$ well separated from the $2\Delta_{CDW}$ threshold for density-like excitations. 
The computation of the Raman kernel for the amplitudon has been detailed in Refs \onlinecite{cea_cdw_prb14,grasset_prb18}, and we report here the main results. Below $T_{CDW}$ the non-linear optical kernel $K(\omega)$ in the symmetric Raman channel can be schematically written as
\begin{equation}
\lb{chi}
K(\omega)= \frac{R_{eff}(T,\Delta_{CDW})}{\omega^2-\omega_0^2-\Sigma(\omega)},
\end{equation}
where the prefactor $R_{eff}\sim \Delta_{CDW}^2$  measures the Raman visibility of the CDW phonon and grows below $T_{CDW}$ proportionally to the CDW order parameter, while  $\Sigma(\omega)$ denotes the (complex) self-energy due to the coupling between the phonon and the Higgs fluctuations $\delta \Delta_{CDW}$:
\begin{equation}
\lb{sigmaph}
\Sigma(\omega)=2g^2\omega_0 \chi_{CDW}(\Omega),
\end{equation}
where $\chi_{CDW}$ denotes the bare susceptibility for amplitude fluctuations, i.e.\
\be
\chi_{CDW}=\frac{4}{N}\sum_\bk \frac{\xi_\bk^2}{E_\bk [(i\Omega_m)^2-4E_\bk^2]}\tanh(E_\bk/2T).
\ee

\noindent After analytical continuation to real frequencies, the renormalized phonon frequency $\Omega_0$ and the peak width $\Gamma_0$ below $T_{CDW}$ can be obtained by solving the two equations 
\bea
\lb{omegaren}
\Omega_0^2=S'(\Omega_0) \\
\lb{gammaren}
\Gamma_{0}=-S''(\Omega_0),
\eea
where we introduced the quantity $S(\omega)$:
\bea
S(\omega)&\equiv& \omega_0^2+\Sigma(\omega)=\omega_0^2[1+(2g^2/\omega_0)\chi_{CDW}(\Omega_0)]\nn\\
&=&\frac{2g^2\omega_0}{N}\sum_\bk \gamma_\bk^2\frac{\omega^2-4\Delta_{CDW}^2\gamma_\bk^2}{E_\bk [(\omega+i\delta)^2-4E_\bk^2]}\tanh(E_\bk/2T).\nn\\
\lb{newomega}
\eea
The last line has been obtained by using the self-consistent equation \pref{selfc}.  The quantity in the rhs of Eq.\ \pref{newomega} scales as the denominator of Eq.\ \pref{higgs}, which identifies the Higgs mode. In particular when $\gamma_\bk=1$ one can rewrite Eq.\ \pref{omegaren} as
\be
\Omega_0^2\simeq 2g^2\omega_0[\Omega_0^2-4\Delta_{CDW}^2]F(\Omega_0),
\ee
showing explicitly that the renormalized phonon frequency is connected to Higgs fluctuations. As mentioned above, in most CDW systems the energy scale set by the quasiparticle excitations is larger than the phonon energy, so that in the self-consistent Eq.\ \pref{omegaren} one can approximately set $\Omega_0\simeq 0$ on the right-hand side. By taking into account the temperature dependence of $F(\Omega)$ one then finds that $\Omega_0(T=0)$ is much smaller than the bare frequency $\omega_0$\cite{rice_ssc74,cea_cdw_prb14,grasset_prb18}.  Conversely, as $T\rightarrow  T_{CDW}$, $\Omega_0\rightarrow 0$. At $T>T_{CDW}$ the Higgs mode is no more defined, but the phonon is still coupled to electronic charge fluctuations at $\bQ$, so that its self-energy simply scales as the electronic charge susceptibility at $\bQ$:
\be
\lb{omegaup}
\Omega_0^2=\omega_0^2\left[1-\frac{2g^2}{\o_0 N}\sum_\bk \frac{\gamma_\bk^2}{E_\bk}\tanh(E_\bk/2T) \right].
\ee
At $T=T_{CDW}$ the right-hand side of Eq.\ \pref{omegaup} reduces to the self-consistency equation \pref{selfc} with $\D_{CDW}=0$, and one recovers the phonon softening due to the lattice instability.  Below $T_{CDW}$ the imaginary part of the Higgs fluctuations provides also the phonon damping, see Eq.\ \pref{gammaren}. In general a momentum dependence of the CDW gap allows for residual quasiparticle excitations below $2\Delta_{CDW}$ even at low $T$. These in turn provide a decay channel for the CDW phonon increasing its broadening, especially as the temperature approaches $T_{CDW}$, as observed in Raman spectra\cite{tsang_prl76,chiang_prl03,grasset_prl19,degiorgi_pnas13}.

In summary, the non-linear optical kernel \pref{chi}  below $T_{CDW}$ is formally equivalent to the ordinary phonon kernel \pref{ksharp}:
\be
\lb{kapp}
K(\omega)\simeq \frac{\Delta^2_{CDW}(T)}{\omega^2-\Omega_0^2+i\omega\G_0},
\ee
allowing us to compute the pump-probe response along the lines outlined in the previous Sections. 
In Fig.\ \ref{fig-cdw} we show the simulated pump-probe spectra of the CDW phonon at selected temperatures below $T_{CDW}$ for broadband THz pulses. Here we choose parameter values appropriate to reproduce the experimental situation where $2\Delta_{CDW}> \Omega_0$, i.e.\ $g/t=0.27$, $\omega_0/t=0.15$. As one can see, the amplitude of the oscillations is rapidly suppressed already at $T/T_{CDW}\simeq0.8$, due to the reduction of the Raman visibility $R_{eff}\sim \Delta^2_{CDW}$. This is consistent with the experiments of Refs \onlinecite{yusupov_prl08,yusupov_natphys10}, where one needs to go significantly below $T_{CDW}$ to detect appreciable oscillations in the probe signal. At the best of our knowledge, the CDW amplitudon has been seen so far only in tritellurides, by means of light excitations in the visible. As discussed in Sec.\ \ref{phonons} within the context of ordinary phonons, a THz pulse tuned at the right frequency is expected to have a much larger efficiency than a visible light pulse. It would be then highly desirable to test the approach proposed here to excite CDW soft phonon modes via THz pulses, especially in CDW dichalcogenides where the lower transition temperature requires cooling down much more to gain enough Raman visibility to excite the mode via visible light.


\section{Discussion and Conclusions}
In the present work we provide a general theoretical scheme to describe pump-probe experiments by means of a quasi-equilibrium approach. In particular we show that the coherent oscillations of the probe signal as a function of the time delay between the pump and the probe can be linked to the resonant behavior of the non-linear optical kernel, showing that not only phonons but any Raman-active collective electronic excitation can be excited by a light pulse via an impulsive-stimulated Raman process. For the sake of concreteness we focus here on the experiments in transmission geometry, and we derive explicitly the transmitted probe field as a function of the observation time $t_g$ and the pump-probe delay $t_{pp}$. We show that the power spectrum of the differential probe signal can be linked to the product of the non-linear kernel $K(\omega)$ times the power spectrum of the squared pump pulse:
\be
\lb{general}
\d E_{probe}(\omega) =K(\omega)\bar A^2_{pump} (\omega).
\ee
This simple relation, along with the existence of a pronunced resonance of the kernel $K(\omega)$ at a certain frequency $\omega_{res}$, represents the starting point to understand the difference between several experimental set up, and the link between pump-probe spectrosocpy and third-harmonic generation. For a visible light pulses only the $\omega\simeq 0$ component of $\bar A^2(\omega)$ is relevant, showing that the mode is excited via a difference-frequency process. For a THz pulse the $\omega\simeq 2\Omega_{pump}$ component of $\bar A^2(\omega)$ is relevant, and the mode is excited via a sum-frequency process. As a consequence, we clearly establish that for ultrafast THz spectroscopy not only the pump pulse duration (which determine the adiabatic vs anti-adiabatic condition) but also its central frequency cooperatively determine the generation of coherent oscillations. By taking advantage of the relation \pref{general}, we discuss the difference between broadband and narrowband THz pulses, explaining why in the former case one generates oscillations at the resonance frequency $\omega_{res}$, while in the latter oscillations occur at twice the pump frequency.

We then discuss three applications of the general approach based on Eq.\ \pref{general}. The first one is the case of Raman-active phonons. We show the analogies and differences between our scheme and the approach based on the equation of motions for the displacement phononic field, and we recast in our language some interesting recent results obtained for THz-stimulated phononic oscillations in insulators\cite{kampfrath_prl17,maehrlein_prb18,johnson_prl19}. 

The second application concerns coherent oscillations of SC collective modes. 
The main advantage of the general relation \pref{general} is that the computation of  the non-linear optical kernel $K(\omega)$ within a specific and controlled approximation allows one to address separately the various physical processes relevant for each system, and to test the results against the experiments. 
This approach can lead to a considerable advantage in the case of collective electronic modes across a phase transition, when the full numerical solution of a time-dependent problem is computationally challenging\cite{spivak_prl04,werner_prb16,millis_prb17,lorenzana_prb18,axt_prb07,axt_prb08,carr_sust13,manske_prb14, manske_natcomm16,kollath_prl17}, and does not always allow one to disentangle the contributions from different excitations channels. This is exactly what happens for the $2\Delta$ oscillations in a superconductor, that have been  ascribed so far either to the excitation of  density-like fluctuations\cite{carbone_pnas12} or to the Higgs mode\cite{shimano_prl12,shimano_prl13,shimano_science14}. While in a generic lattice model the Higgs is always  excited by a light pulse \cite{axt_prb07,axt_prb08,carr_sust13,manske_prb14, manske_natcomm16,kollath_prl17}, its contribution to the differential probe field \pref{general} is crucially determined by its Raman visibility. We then show that in general the experimental observation of THz-induced coherent oscillations at twice the gap value\cite{shimano_prl13} or at twice the pump frequency\cite{shimano_science14} could be well reproduced by considering only density-like fluctuations, which certainly dominate the kernel $K(\omega)$ in the clean limit, as it has been widely discussed within the context of the third-harmonic generation\cite{cea_prb16,aoki_prb16,cea_leggett_prb16,shimano_prb17,cea_prb18,shimano_cm19,silaev_cm19}. More recently, it has been pointed out\cite{shimano_cm19,silaev_cm19}  that in the presence of disorder the hierarchy of the two  contributions can be reverted, making the Higgs contribution to the non-linear kernel $K(\omega)$ the dominant one. The main reason is that disorder triggers paramagnetic-like processes able to enhance the Raman-like visibility of the Higgs mode, which then rapidly overcomes the density-like diamagnetic response. A somehow counterintuitive result of these calculations is that
an extremely small amount of disorder is enough to make  paramagnetic-like processes dominant. However, if this is the case the Leggett mode, recently observed by pump-probe spectroscopy in the  multiband  MgB$_2$ superconductor, should not be visible, since its Raman visibility is strongly constrained to the relevance of diamagnetic-like processes\cite{cea_leggett_prb16,chubukov_raman_prb17}. So far, the experimental observations of Ref.\ \onlinecite{giorgianni_natphys19} have been very well reproduced by the theoretical scheme based on Eq.\ \pref{general}, with a calculation of the non-linear kernel done in the clean limit, so more work is needed to reconcile these results with the outcomes of Refs \onlinecite{shimano_cm19,silaev_cm19}. In addition, the comparison between pump-probe experiments and THG in a system where multiple resonant excitations  appear in the non-linear kernel $K(\omega)$ requires some additional care on the matching conditions for the sum-frequency process. In the case of MgB$_2$ one has in principle three excitations, corresponding to twice the two gap values $\Delta_{1,2}$ and to the Leggett $\omega_L$, with $2\Delta_1<\omega_L<2\Delta_2$. When the system is excited with a narrowband pulse with central frequency $\Omega_{pump}$ the only collective excitation which contributes to the non-linear optical response is the one which matches   the sum-frequency condition. In Ref.\ \onlinecite{shimano_cm19} it has been argued that the Leggett mode gives a negligible contribution to the THG when the system is excited with a light pulse matching the larger gap value, $2\Omega_{pump}\simeq 2\Delta_2\gg \omega_L$. However, this result cannot be simply ascribed to the effect of disorder, since even in the clean case the frequency of the pump would be too high to excite the Leggett mode, and it will then always give negligible contribution to the non-linear kernel. A more compelling comparison could be the THG signal obtained for two pump frequencies, one matching the Leggett and one matching twice the larger gap. Finally, it is worth noting that calculations in Refs \onlinecite{shimano_cm19,silaev_cm19} are done in the Born limit of weak disorder, where the Higgs resonance stays untouched and disorder only affects the Raman visibility of the Higgs. However, as disorder increases it has been shown that the mixing between the SC phase and amplitude modes spoils the spectral weight of the Higgs fluctuations at $2\Delta$\cite{cea_prl15}. It would be then very interesting to study experimentally the fate of the $2\Delta$ oscillations as disorder increases in conventional superconductors as NbN. 

Finally, we discussed the case of the Higgs mode in a CDW transition. In this case the amplitude fluctuations of the CDW order parameter couple to the CDW phonon giving rise to a soft "amplitudon" phonon mode at a frequency $\Omega_0$ much lower than twice the gap value $2\Delta_{CDW}$, allowing one to spectroscopically disentangle the two contributions that appear at the same scale $2\Delta$ in a superconductor. We have shown that in this case the pump-probe response looks pretty much the same of an ordinary phonon, with remarkable differences due to the temperature dependence both of the phonon frequency $\Omega_0$ and of its Raman visibility. These findings are consistent with the observation of the amplitude CDW mode in tritellurides via visible-light excitations\cite{yusupov_prl08,yusupov_natphys10}. However, for CDW systems a THz-pump counterpart of these experiments is still lacking, both for tritellurides and for the other class of CDW materials based on dichalcogenides. Our general scheme, based on the prediction \pref{general} for the form of the power spectrum of the differential probe field, could serve as a guideline to design these experiments. It can also trigger further theoretical work aimed at a quantitative computation of the non-linear optical kernel in CDW materials, with the ultimate goal to use pump-probe protocols to gain crucial informations on the microscopic mechanisms responsible for the CDW transition in these systems.

\acknowledgements
We acknowledge financial support by Italian MAECI under the Italian-India
collaborative  project  SUPERTOP-PGR04879,  and by the European Commission under the Graphene Flagship, contract CNECTICT-604391. 

\vspace{1cm}

\appendix
\section{Long-time decay of the transmitted pulse}
\label{longtime}

Here we derive analytically the asymptotic behavior of the non-linear optical kernel in the case of an approximated parabolic band. We are interested in the function:
\be
K(t)\equiv \int_{-\infty}^{\infty}\frac{\,d\omega}{2\pi}K(\omega)e^{-i\omega t}
\ee
in the limit $\Delta t\gg 1$. Using the reality of $K(t)$, the previous integral can also be written as
\be\lb{Kt}
K(t)= \int_{0}^{\infty}\frac{\,d\omega}{\pi}\mathrm{Re}\left\{K(\omega)e^{-i\omega t}\right\},
\ee
which allows us to consider positive frequencies $\o>0$ only.

\begin{figure}[htb]
\centering
\includegraphics[scale=0.35]{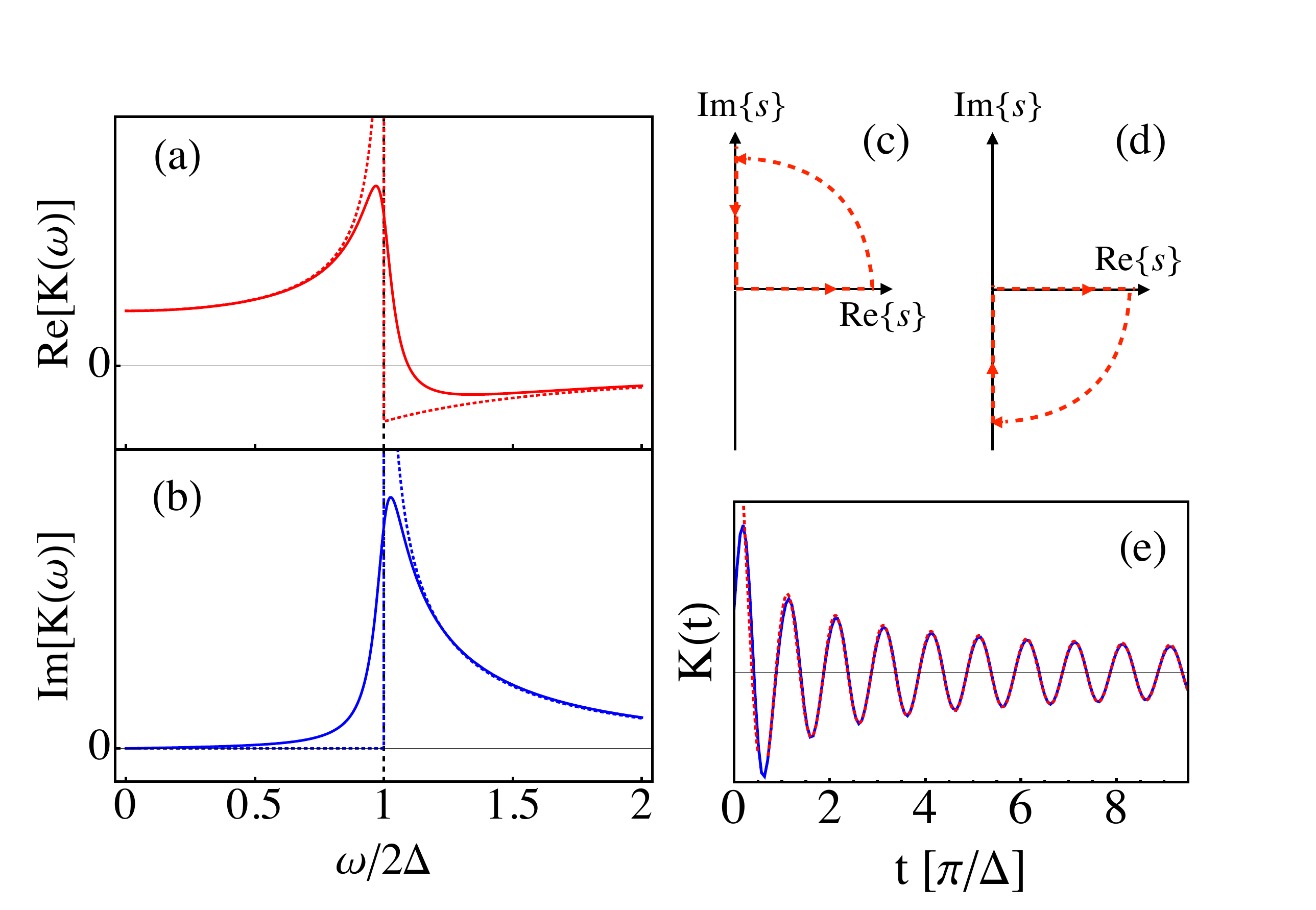}
\caption{
Real (a) and imaginary (b) part of the kernel $K(\omega)$ as a function of $\omega/2\Delta$. The continuous line corresponds to the finite broadening $\gamma=0.1\Delta$ while the dotted one to $\gamma=0$. Here we used $\omega_D=10\Delta$, that corresponds to the relatively weak dimensionaless SC coupling $\lambda_{SC}\simeq0.33$. (c)-(d) Projection of the integration path of the integrals $\int_0^\infty \,ds\frac{e^{\pm is}}{\sqrt{s}}$ on the imaginary axis. The $+$ or $-$ signs appearing in the exponent of the integrand require to choose the positive (c) or negative (d) imaginary axis, respectively. (e) Comparison between the function $K(t)$ (continuum blue line), as obtained by the numerical integration of the Eq.\ \pref{Kt}, and the long-time estimate of Eq.\ \pref{Kt3} (dotted red line).
}\label{kernel}
\end{figure}

For the sake of simplicity we consider the illustrative case of weak SC coupling $\omega_D\gg\Delta$ and zero spectral broadening $\gamma=0$, so that the Eq.\ \pref{KconstDOS} reduces to:
\be
K(\omega)= -\frac{2\Delta^2N_F}{m^2}\frac{\tan^{-1}\left(\frac{\o}{\sqrt{4\Delta^2-\o^2}}\right)}{\o\sqrt{4\Delta^2-\o^2}}.
\ee
Notice that while $K(\omega)$ is purely real for $\omega^2<4\Delta^2$, for $\omega^2>4\Delta^2$ a finite imaginary part appears, see Fig.\ \ref{kernel}a-b. By distinguishing the two cases we obtain:
\begin{widetext}
\be
K(\omega)=-\frac{\Delta^2N_F}{m^2}\left\{
\Theta(4\Delta^2-\omega^2)2\frac{\tan^{-1}\left(\frac{\o}{\sqrt{4\Delta^2-\o^2}}\right)}{\o\sqrt{4\Delta^2-\o^2}}+
\Theta(\omega^2-4\Delta^2)\left[
\frac{\ln\left(\frac{\o-\sqrt{\o^2-4\Delta^2}}{\o+\sqrt{\o^2-4\Delta^2}}\right)}{\o\sqrt{\o^2-4\Delta^2}}+
\frac{i\pi}{\o\sqrt{\o^2-4\Delta^2}}
\right]
\right\},
\ee
\end{widetext}
where the branch of the complex square-root has been chosen in order to satisfy $\sqrt{4\Delta^2-\o^2}=-i\sqrt{\o^2-4\Delta^2}$ $\forall$ $\o>2\Delta$, that is imposed by requiring $K(\o)$ to be a retarder Green's function, obtained as analytical projection from the upper half plane, i.e. $
\o\equiv\lim_{\g\to0^+}\o+i\g$.

In the long time limit the integral of the Eq.\ \pref{Kt} is dominated by the singular contribution, $\o^2\sim4\Delta^2$, of $K(\omega)$, that can be written as:
\be\lb{Ksing}
K_{sing}(\omega)=-\frac{N_F}{4m^2}\frac{\pi\sqrt{\Delta}}{\sqrt{|\o-2\Delta|}}\left[\Theta(2\Delta-\omega)+i\Theta(\omega-2\Delta)\right].
\ee
Substituting the Eq. \pref{Ksing} into \pref{Kt} gives:
\bea
\lb{Kt2}
&K(t)\sim -\frac{N_F}{4m^2} \mathrm{Re}\left\{
\int_{0}^{2\Delta}\,d\o\frac{e^{-i\o t}}{\sqrt{2\Delta-\o}}+i\int_{2\Delta}^\infty\,d\o\frac{e^{-i\o t}}{\sqrt{\o-2\Delta}}\right\}\nn\\
&\sim -\frac{N_F}{4m^2} \sqrt{\frac{\Delta}{t}}\mathrm{Re}\left\{e^{-i2\Delta t}\left[\int_0^{\infty}\,ds \frac{e^{is}}{\sqrt{s}}+i\int_0^{\infty}\,ds \frac{e^{-is}}{\sqrt{s}}\right]
\right\}\nn\\
\eea
where we introduced the integration variable $s=\o t$ and we put $2\Delta t=\infty$ in the first integral of the rhs. The two integrals 
$$
\int_0^\infty \,ds\frac{e^{\pm is}}{\sqrt{s}}
$$
can be easily computed by projecting the path of integration on the positive ($+$) or negative ($-$) imaginary axis (see the Fig.\ \ref{kernel}c-d), which leads to:
$$
\int_0^\infty \,ds\frac{e^{\pm is}}{\sqrt{s}}=e^{\pm i\pi/4}\int_0^\infty \,ds\frac{e^{-s}}{\sqrt{s}}=e^{\pm i\pi/4}\sqrt{\pi}.
$$
 The Eq. \pref{Kt2} then finally gives:
\be\lb{Kt3}
K(t)\sim -\frac{N_F}{2m^2} \sqrt{\frac{\pi\Delta}{t}}\cos\left(2\Delta t-\frac{\pi}{4}\right).
\ee
\\
Although the Eq.\ \pref{Kt3} has been obtained in the long time limit, it provides a quite precise estimate of the function $K(t)$ up to very short time scales, as we show in the Fig.\ \ref{kernel}, where the continuum line represents the result of the numerical integration of the Eq.\ \pref{Kt}, while the dotted one represents the estimate \pref{Kt3}.

\bibliography{Literature.bib}

\end{document}